\RequirePackage{fix-cm}
\documentclass[smallextended]{svjour3}       
\smartqed  
\usepackage{graphicx}
\usepackage{url}
\usepackage{xspace}
\usepackage{color}
\usepackage{tcolorbox}
\usepackage{color, colortbl}
\usepackage{mathtools}
\usepackage{pifont}
\definecolor{LightCyan}{rgb}{0.88,1,1}

\newif\ifdraft
\drafttrue

\ifdraft
  \newcommand{\grex}[1]{{\color{blue}\emph{Gregorio says: #1}}\xspace}
  \newcommand{\andrea}[1]{{\color{red}\emph{Andrea says: #1}}\xspace}
  \newcommand{\jgb}[1]{{\color{red}\emph{Jesus says: #1}}\xspace}
  \newcommand{\bjorn}[1]{{\color{red}\emph{Björn says: #1}}\xspace}
  \newcommand{\jonas}[1]{{\color{red}\emph{Jonas says: #1}}\xspace}
  \newcommand{\cn}{~\textcolor{violet}{[citation needed]}}
  \newcommand{\fixme}[1]{{\textcolor{red}{[FIXME] #1}}\xspace}
  \newcommand{\original}[1]{{\color{brown}\emph{{\bf From the original paper:} #1}}\xspace}
\else
  \usepackage[disable]{todonotes}
  \newcommand{\grex}[1]{}
  \newcommand{\andrea}[1]{}
  \newcommand{\jgb}[1]{}
  \newcommand{\bjorn}[1]{}
  \newcommand{\jonas}[1]{}
  \newcommand{\fixme}[1]{}
  \newcommand{\cn}{}
  \newcommand{\original}{}  
\fi

\usepackage[inline]{enumitem}

\newcommand{\tbd}{{\color{red}\emph{To be done.}\xspace}}

{\begin{center}\vspace{1mm}\noindent\begin{Sbox}\begin{minipage}{0.95\columnwidth}}%
{\end{minipage}\end{Sbox}\fbox{\TheSbox}\end{center}\vspace{1mm}}

\definecolor{goldenyellow}{rgb}{1.0, 0.87, 0.0}

\definecolor{1.00}{rgb}{0.41, 0.67, 0.42} 
\definecolor{0.99}{rgb}{0.44, 0.70, 0.46} 
\definecolor{0.98}{rgb}{0.47, 0.73, 0.5} 
\definecolor{0.97}{rgb}{0.50, 0.76, 0.55} 
\definecolor{0.96}{rgb}{0.53, 0.79, 0.64} 
\definecolor{0.95}{rgb}{0.56, 0.82, 0.72} 
\definecolor{0.94}{rgb}{0.59, 0.85, 0.77} 
\definecolor{0.93}{rgb}{0.62, 0.88, 0.83} 
\definecolor{0.92}{rgb}{0.65, 0.91, 0.87} 
\definecolor{0.91}{rgb}{0.68, 0.94, 0.92} 
\definecolor{0.90}{rgb}{0.71, 0.97, 0.95} 
\definecolor{0.89}{rgb}{0.74, 1.0, 1.0} 
\definecolor{0.88}{rgb}{0.77, 1.0, 1.0} 
\definecolor{0.87}{rgb}{0.80, 1.0, 1.0} 
\definecolor{0.86}{rgb}{0.83, 1.0, 1.0} 
\definecolor{0.85}{rgb}{0.86, 1.0, 1.0} 
\definecolor{0.84}{rgb}{0.89, 1.0, 1.0} 
\definecolor{0.83}{rgb}{0.92, 1.0, 1.0} 
\definecolor{0.82}{rgb}{0.95, 1.0, 1.0} 
\definecolor{0.81}{rgb}{0.98, 1.0, 1.0} 
\definecolor{0.80}{rgb}{1.00, 1.0, 1.0}

\begin{document}

\title{Development Effort Estimation in Free/Open Source Software from Activity in Version Control Systems
}

\titlerunning{Development Effort Estimation from Activity in Version Control Systems}        

\author{Gregorio Robles         \and
        Andrea Capiluppi \and 
        Jesus M. Gonzalez-Barahona \and
        Bj\"orn Lundell \and
        Jonas Gamalielsson
}


\institute{G. Robles, J.M. Gonzalez-Barahona \at
              Department of Telematic and Computational Systems Engineering \\
              Universidad Rey Juan Carlos, Spain \\
              \email{gregorio.robles@urjc.es, jesus.gonzalez.barahona@urjc.es}
           \and
           A. Capiluppi \at
              Department of Computer Science \\
              University of Groningen, The Netherlands \\
              \email{a.capiluppi@rug.nl}
           \and
           B. Lundell, J. Gamalielsson \at
              School of Informatics \\
              University of Sk\"ovde, Sweden \\
              \email{bjorn.lundell@his.se, jonas.gamalielsson@his.se}
}

\date{Received: date / Accepted: date}

\maketitle

\begin{abstract}
Effort estimation models are a fundamental tool in software management, and used as a forecast for resources, constraints and costs associated to software development. For Free/Open Source Software (FOSS) projects, effort estimation is especially complex: professional developers work alongside occasional, volunteer developers, so the overall effort (in person-months) becomes non-trivial to determine.

The objective of this work it to develop a simple effort estimation model for FOSS projects, based on the historic data of developers' effort. The model is fed with direct developer feedback to ensure its accuracy.

After extracting the personal development profiles of several thousands of developers from 6 large FOSS projects, we asked them to fill in a questionnaire to determine if they should be considered as full-time developers in the project that they work in. Their feedback was used to fine-tune the value of an effort threshold, above which developers can be considered as full-time.

With the help of the over 1,000 questionnaires received, we were able to determine, for every project in our sample, the threshold of commits that separates full-time from non-full-time developers.
We finally offer guidelines and a tool to apply our model to FOSS projects that use a version control system.

\end{abstract}

\section{Introduction}
\label{sec:intro}

Effort estimation models are invaluable tools in software management: they help gaining insights on past resources and associated costs; and they serve as models to forecast resource demand and allocation, as well as dealing with predicted constraints. Traditionally, effort estimation is generally used by companies in the earlier stages of a software project to estimate the number of developers, and the amount of time that it will require to develop a software. 

In the context of Free/Open Source Software (FOSS) development, there has hardly been a formal way of tracking the effort of developers, especially in its original volunteering form. It was argued that, in some circumstances, small-team development could eventually evolve~\cite{capiluppi2007cathedral} into a distributed development (e.g., `bazaar'). In such cases, effort is even more complex to track, since the level of contributions varies along an onion-type model~\cite{crowston2005social}, where the outer the layer the smaller the code contribution, but the larger the user base, and the bug reporting facility. The inner the layer, the more pronounced the development effort.

This volunteer-based model for FOSS development was later complicated by the participation of companies, alongside volunteers~\cite{riehle2014paid}. Companies devoted staff and effort into a FOSS project to enter a saturated market, or to gain a foothold in a specialised development, without deploying vast amounts of resources and recreate a competitor product from scratch~\cite{fitzgerald2006transformation}. This hybrid model introduces further complexity in the effort estimation modeling: if the volunteers can still be modeled alongside the basic onion-model, the involvement of external companies and their focused effort become quite challenging to describe and model. 

The effort estimation model presented depends on the threshold value of a FOSS project, and on the analysis of its publicly available data on version control systems (VCSs), such as git. Other projects can easily create their tailored model, and survey their developers to obtain their threshold value. As a further contribution of this work, we propose the threshold values of six large FOSS projects. These values can be re-used, compared, or validated with other projects. 

Given the fluid nature of FOSS development, all the existing approaches to FOSS effort estimation are currently missing two pieces of crucial information. The first is a precise measure of past effort of the developers, even in the more sophisticated models. In traditional effort estimation models, software projects are able to estimate future effort, because they know their past effort. Given the increasing involvement of companies in FOSS development, this aspect is fundamental to (i) allow potential new participant companies to evaluate the current developers effort, and (ii) fine tune their participatory effort in the FOSS projects.

The second piece of crucial information that is missing is some sort of validation by the developers themselves. Assigning developers to the wrong tiers, or using a too wide (or too small) time window to evaluate the commits activity, produces different values of estimated effort.

For this purpose, in this paper we surveyed over 1,000 developers working on 6 large FOSS projects, with the aim to count how many developers consider themselves as `\textit{full-timers}' in each project. The number of commits (in a six months' timespan) of these self-identified full-time developers was later used as the \textit{threshold $\theta$} that minimises the error to differentiate full-timers from other contributors. We used that threshold  $\theta$ to estimate the effort produced by developers in each of the analysed FOSS projects. The contribution of this work is, to the knowledge of the authors, the first developer-validated model to FOSS effort estimation available in literature.

The present work is an extension of a previously published paper~\cite{robles2014estimating}. Its \textit{Future Work} section posited that: ``\textit{We envisage to expand this study by 1) studying other FOSS projects to ascertain if our method is applicable in general, and if it is to what extent; 2) performing a scientific experiment to obtain margins of error for the estimation of error for full-time developers (...)}''.

 The current paper should therefore be considered as the enactment of the future work proposed in~\cite{robles2014estimating},
 and provides the following specific contributions:
\begin{itemize}
  \item We provide analysis and results for five more industrial FOSS projects.
  \item We offer an extension of a previous SLR on effort estimation including papers from 2016-2020.
  \item We clarify the model, offering an example of how it can be used in practice.
  \item We further develop on the concepts of goodness and compensation
  \item We discuss about the estimation error of the model.
  \item We discuss the repercussions for practitioners.
\end{itemize}

This paper is articulated as follows: Section~\ref{sec:_vision} introduces the vision of this research. Section~\ref{sec:_related} deals with related work, especially focusing on the effort estimation models built for the various generations of FOSS (e.g., fully volunteered projects, hybrid FOSS projects, company-led FOSS projects). Section~\ref{sec:model} introduces the model for effort estimation that we propose, alongside the terms and definitions we used throughout this study; Section~\ref{sec:projects} introduces the characteristics of the analysed projects, while Section~\ref{sec:experiment} describes the survey that was circulated to the developers of the FOSS projects, and how the threshold $\theta$ was evaluated for each project. Section~\ref{sec:results} shows how the model is deployed for the OpenStack project, while Section~\ref{sec:_results_replicated} proposes further replication, when considering five other projects. Section~\ref{sec:discussion} discusses the findings, especially considering what sort of benefits we envisage for FOSS communities and companies. Section~\ref{sec:conclusions} finally concludes.

\section{Vision of the research}
\label{sec:_vision}
\begin{figure*}[htpb]
\begin{center}
  \includegraphics[keepaspectratio=true,width=0.75\textwidth]{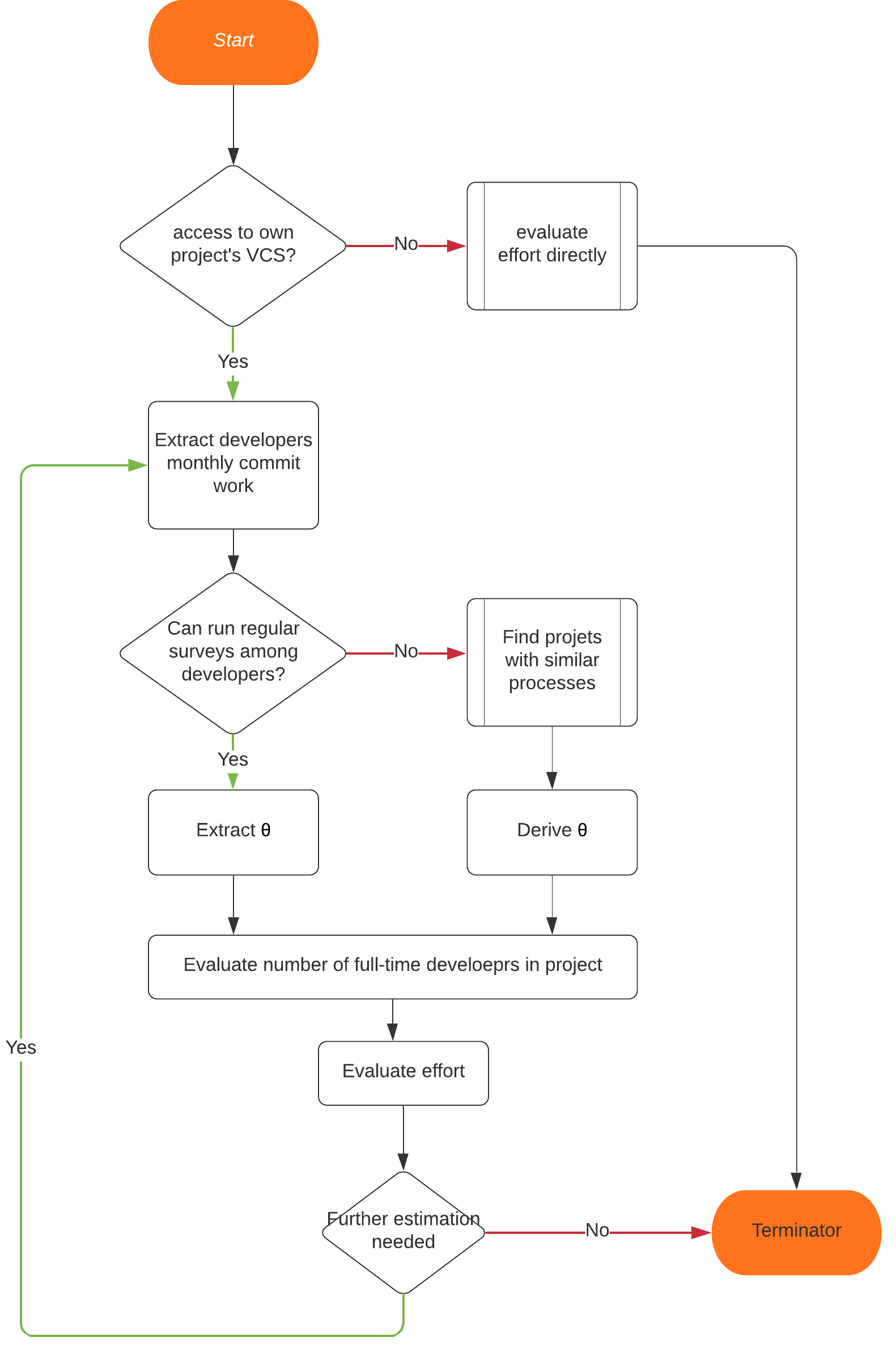}
\label{fig:_vision}
\caption{Flowchart to run our effort estimation model in other FOSS projects.}
\end{center}
\end{figure*}

The model that we propose below is based on an overarching vision, that is set to maximise the applicability and reproducibility of our model in other FOSS projects. As depicted in the flowchart of Figure~\ref{fig:_vision}, the model requires the parsing of a project's VCS, and that is a relatively simple resource to have an access to, with modern software engineering tooling. The model then uses the responses of developers to a survey to establish a value of commits ($\theta$) above which developers should be considered as full-time. 

The second data source of our effort estimation model (i.e., creating, distributing, collecting and analysing the developer surveys) can be time consuming and requires commitment from a FOSS project's management board. To combat this, our model has been built also for those projects that cannot (or do not want) to directly and regularly survey their developers: the vision of this research is to build enough evidence from several FOSS projects, and to obtain threshold values for \textit{classes} of similar projects. Those shared thresholds can then be reused for other projects, thus removing the need to regularly survey developers.

\section{Related Research}
\label{sec:_related}

Traditional effort estimation models are generally used by organizations in the early stages of a project to estimate total effort, number of developers and time (e.g., duration) needed to develop the software. Effort estimation in traditional commercial software has been approached by means of models, either top-down (e.g., by analogy or by expert judgement) or bottom up (e.g., via regression models using past data). In the area of effort estimation, a comprehensive and systematic literature review on effort estimation was presented  in 2007 in~\cite{jorgensen2007systematic}. The most evident result was to observe that over half of the papers surveyed are based on \textit{history-based} evaluations to build models to estimate future effort.

\begin{figure*}[htpb]
\begin{center}
  \includegraphics[keepaspectratio=true,width=0.7\textwidth]{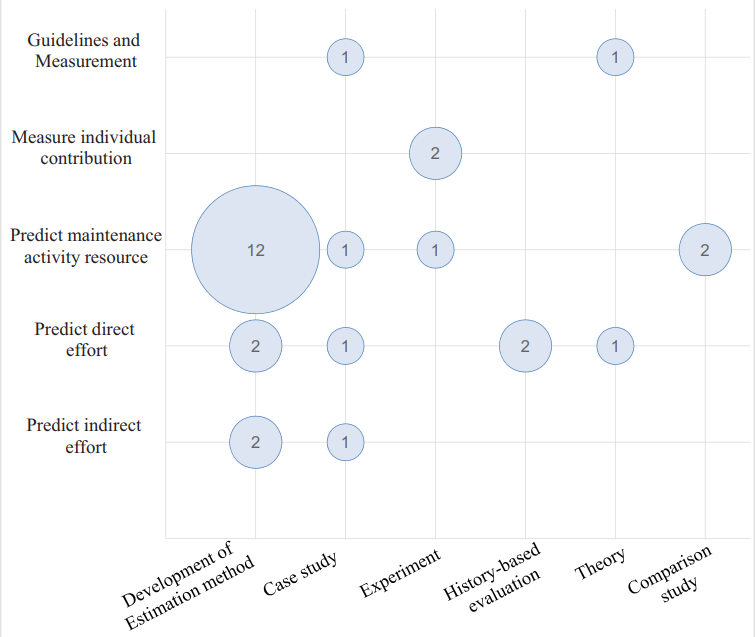}
  \vspace{1cm}
  \includegraphics[keepaspectratio=true,width=0.7\textwidth]{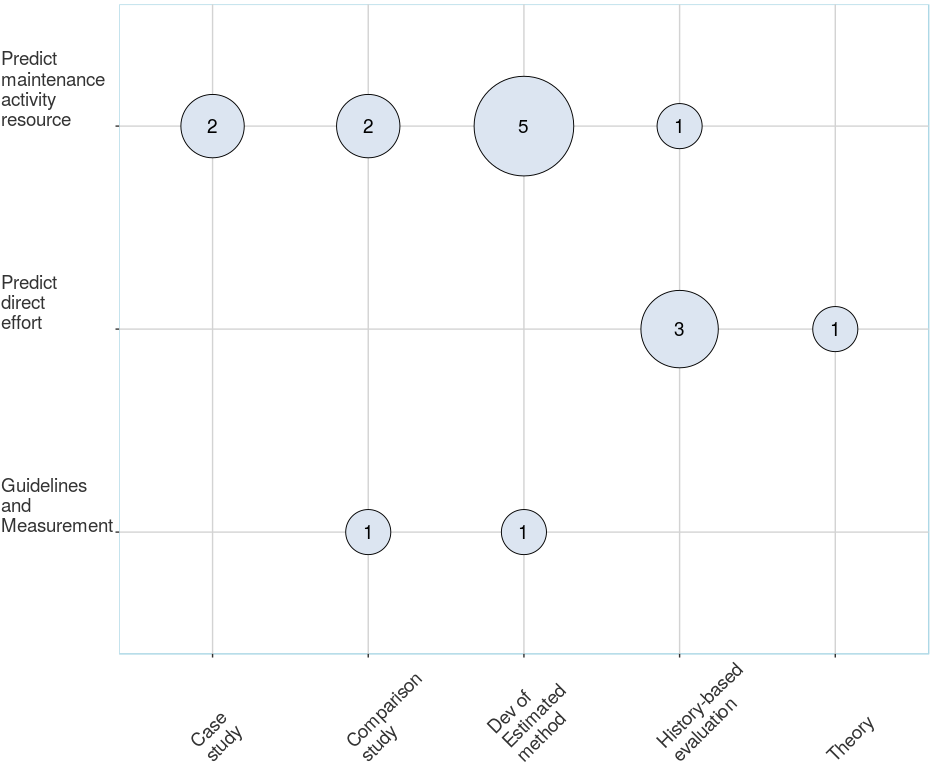}
	\caption{Topic of the SLR conducted in 2016 by~\cite{wu2016maintenance} (top); additional SLR performed for this study (bottom)}
	\label{fig:SLR_add}
\end{center}
\end{figure*}

Similarly to the systematic literature review (SLR) of 2007~\cite{jorgensen2007systematic}, the research works around effort estimation (specifically for FOSS systems) were extensively analysed in 2016 by Wu et al.~\cite{wu2016maintenance}. The type of study proposed by the authors, alongside the object of the predictions, were analysed and clustered, as visible in the top part of Figure~\ref{fig:SLR_add}. Different types of studies were identified, based on the \textit{goal} of the relative research (`Guidelines and Measurement', `Measure individual contribution', `Predict maintenance activity resource', `Predict direct effort' and `Predict indirect effort') and the \textit{approach} used in the research (`Development of estimation method', `Case study', `Experiment', `History-based evaluation', `Theory' and `Comparison study'). The largest set of research studies was found to be under the `Development of estimation method' to `predict maintenance activity resources' (comparably to~\cite{jorgensen2007systematic}).

In order to include the more recent literature, we repeated the process presented in the 2016 SLR. We analysed the papers that were written between 2016 and 2020, using the same search string\footnote{As per the original paper: (“Open source” OR OSS OR FOSS OR FLOSS OR opensource OR libre OR free) AND (maintain OR maintenance OR evolve OR evolution OR “bug fix*” OR “bug-fixing” OR “defect fixing” OR “defect correction” OR “defect resolution” OR effort OR cost OR estimat*, predict*) AND (empirical OR validation OR experiment OR evaluation OR “case study”).}, and the same search engines\footnote{Inspec, Compendex, IEEE Xplore, ACM Digital Library, Science Direct and Google Scholar.}. We obtained 2,279 candidate studies from the database search (without duplicates). After reading abstract and title, we obtained 41 articles; after reading each, we gathered 16 studies that comply with the search criteria from the original SLR. The labels of the SLR were applied to these papers, as visible in the bottom part of Figure~\ref{fig:SLR_add}. 

Similarly to the original study, the largest category was found again to be the studies that predict maintenance activity resources: although most of the papers design and/or deploy a new estimation method (\cite{honel2018changeset,mi2016empirical,thung2016automatic,zhao2016discussions}, mostly focused on bug-fixing efforts), we also found three studies that used history-based evaluations to directly predict effort~\cite{malhotra2020using,porru2016estimating,yang2016empirical}.

\paragraph{Applicability of Traditional Effort Estimation techniques to FOSS}
The FOSS paradigm, at least in its original form, was deemed as difficult, if not impossible, to fit under the most well-known estimation models (for instance the COCOMO I or COCOMO II models~\cite{barry1981software,boehm2000software}). Various research attempts have showed that fundamental assumptions and constructs could not be applied to the FOSS domain; whereas other aspects were simply absent from commercial organizations~\cite{amor2006effort,capiluppi2013effort,kalliamvakou2009measuring,robles2014estimating}.

More specifically, Amor \textit{et al.}~\cite{amor2006effort} proposed to measure the \textit{total} effort invested in a project by characterizing \emph{all} the activities of developers that can be traced: commits, e-mails, bug activities, among others. Kalliamvakou \textit{et al.}~\cite{kalliamvakou2009measuring} replicated the study and included the contribution of developer effort. 

Capiluppi et al.~\cite{capiluppi2013effort} determined the average hours worked per day by the Linux kernel developers. They characterized the Linux kernel by the time of day when the commits were observed in the repository and when the author worked most frequently. They divided a working day into traditional office hours (from 9am to 5pm), after office hours (5pm to 1am) and late hours (from 1am to 8am). The authors found that within the Linux kernel community the effort was constant throughout the week, which suggested the need for different estimation models from the ones traditionally used in industrial settings, where the work schedule is presumed to be 9am-5pm, Monday to Friday.

A more general work to address the commit frequency was presented in~\cite{kolassa2013empirical}. Using the \url{Ohloh.net} repository, the authors evaluated the overall commit frequency, and the average number of commits that developers featured on. The same repository was also analysed to evaluate the size of a typical commit~\cite{kolassa2013model}, where the authors found power laws underlying the patterns of commit size. Later on, the frequency of commits was analysed in~\cite{hou2014empirical} for two Apache projects, POI and Tomcat. The authors also found that commits follow power laws, with large bursts and long tails. Power laws were again found in 4 Apache projects, when analysing the time interval between commits and files committed~\cite{ma2014dynamics}.


Mockus et al.~\cite{mockus2000identifying} showed that, out of nearly 400 programmers in the Apache project, the 15 most productive ones contributed 88 percent of the total lines of code (LOC). They compared those 15 Apache developers with programmers in five other commercial projects. They defined \textit{code productivity} as the mean number of lines of code per developer per year (KLOC/developer/year). Koch~\cite{koch2002effort,koch2008effort} reported that in FOSS projects, the distribution of effort between participants (programmers) is skewed as in Mockus et al.~\cite{mockus2002two}. Indeed, the majority of code was written and most of the effort spent by just a few contributors. This paper presents and illustrates the use of a similar approach, when identifying who these `major contributors' are, but instead of imposing an \textit{a-priori} threshold above which we consider a developer to be a `super' or a `major' contributor (or even a \textit{hero}~\cite{agrawal2018we}), we validate that threshold with the help of the developer surveys. To the best of our knowledge, the identification of developer types in this way is novel.

Moulla et al.~\cite{moulla2013cocomo,moulla2014application} applied the COCOMO model to the TRIADE (version 7a) FOSS project, using LOCs and regression models with COSMIC Function Points as the independent variable~\cite{dumke2016cosmic,abran20163}. They reported that in terms of effort, development of software based on FOSS has advantages over development from scratch. Fernandez-Ramil et al.~\cite{fernandez2009does} used linear regression models to estimate effort and duration for FOSS projects with LOC as the independent variable. Yu~\cite{yu2006indirectly} also used linear regression models with LOC as the independent variable to indirectly predict maintenance effort in free and FOSS projects. Effort was measured by examining the LOC added, deleted and modified. Anbalagan et al.~\cite{anbalagan2009predicting} investigated predicting how much time developers spend in corrective maintenance of FOSS projects. Their study focused on 72,482 bug reports from over nine releases of Ubuntu, a Linux distribution system. They used a linear regression model with `bug reports corrected by developer’ as the independent variable to predict the time taken to correct bugs. They observed that for FOSS estimation maintenance effort is lower than for proprietary development due to higher code quality. 

According to Capra et al.~\cite{capra2007economics,capra2008empirical,capra2010economics}, FOSS shows a slower growth of maintenance effort over time. Studies by Koch et al.~\cite{koch2002effort,koch2004profiling,koch2008effort} on effort modelling and developer participation in FOSS projects show that the number of participants, other than programmers, was about one order of magnitude larger than the number of programmers. In summary, a number of studies are available on software effort estimation but few discuss effort and duration in FOSS projects using linear regression models. The categorization of contributors and the size and choice of the datasets also present a challenge for research on FOSS projects estimation. 

\begin{table*}[htpb!]
\begin{tabular}{p{1.5cm}lp{1.8cm}p{4cm}p{1.85cm}}
                     & \textbf{Ref}  & \textbf{Type}              & \textbf{Attributes}                                              & \textbf{Validated w developers} \\\hline
Source code          & \cite{yu2006indirectly}   & INDIRECT          & lag time, source code changes                           & NO                        \\
                     & \cite{alomari2015slicing}    & INDIRECT          & lag time, delta LOCs, delta files                       & NO                        \\
Process based        & \cite{abdelmoez2012bug}    & INDIRECT          & Bugs data                                               & NO                        \\
                     & \cite{ahsan2009program}    & INDIRECT          & Bugs data                                               & NO                        \\\hline
People-based estimation                     & \cite{koch2002effort}   & DIRECT            & Proxy of \{lag time, LOCs\}                             & NO                        \\
                     & \cite{capra2007economics}   & DIRECT            & Proxy of \{lag time, LOCS, methods\}                    & NO                        \\
                     & \cite{koch2008effort}   & DIRECT            & Proxy of \{active developers, lag time, LOCS, methods\} &                           \\

Activity-based
estimation                     & \cite{amor2006effort}    & DIRECT            & Proxy of \{LOCs, commits, emails, bug activity\}        & NO                        \\
                     & \cite{capiluppi2013effort}   & DIRECT            & Proxy of \{active developers, lag time, LOCs\}          & NO       \\\hline
People- \& activity-based & \cite{robles2014estimating}   & DIRECT + INDIRECT   & Proxy of \{commits\}, surveys                                    & YES                       \\
	
\rowcolor{LightCyan} People- \& activity-based & ours & DIRECT + INDIRECT & Proxy of \{commits\}, surveys                                  & YES                       \\
                     &      &                   &                                                         &                           \\\hline

\end{tabular}
\label{tbl:_summary-direct-indirect}
\caption{Summary of past works based on the use of direct and indirect measurements. \colorbox{LightCyan}{Highlighted} are the characteristics of the present paper.} 
\end{table*}

\paragraph{Effort Estimation in hybrid FOSS systems}
The terminology around FOSS started to change as long as volunteer communities began to collaborate with industrial entities. One of the first research studies that identified the possibility of a mixed development approach was the seminal work by Lerner
and Tirole~\cite{lerner2002some}. Various strategies were formulated, where commercial enterprises could be profitable using FOSS as a strategic asset.

A more formal description of the grades of involvement of commercial companies into FOSS projects was formulated in two research works~\cite{shah2006motivation,capra2008empirical}. The first work was focused on the motivations of developers and how that was affected by the type of FOSS development. In the second work, and using `governance' as the term of reference, Capra et al. produced various levels of involvement, from volunteer-based, to company-driven FOSS projects.

\paragraph{Summary of past work}
Previous work on FOSS effort estimation has been typically run by detecting effort using \textit{direct} measurements (e.g., number of developers) and applying proxies to those in order to derive a measurement of effort. In other cases, effort was derived \textit{indirectly}, by using various other measurements, and inferring a relationship to the effort devoted to a software project. Table~\ref{tbl:_summary-direct-indirect} summarises the past works in terms of how direct and indirect measurements (and which ones) were used when deriving effort estimation models. As visible, the approach that we propose in our paper (and that has been trialled in its earlier inception at~\cite{robles2014estimating}) is grounded in both direct and indirect measurements. Furthermore, its results have been validated with developers' responses: this alone is the one aspect that sets our model apart from any other attempt at estimating the effort produced by FOSS developers.

In general, effort estimation models for FOSS projects are based on the data that are collected in the various development activities; these data (especially when they have large variance, or are heterogeneous) need transformation parameters (e.g., logarithmic) that might be needed to normalise a distribution, or to take into account different types of data for the same model. One example of the latter would be devising a transformation parameter to consider all the different activities (emails, blogs, online discussions, etc., as well as the actual code commits) that a developer is engaged into~\cite{amor2006effort}.

This type of models suffers from a couple of issues: first, it is well known that contributions to FOSS projects are not uniform. A few developers are responsible for a major amount of the work, while a large number of developers contribute with a comparatively small one~\cite{koch2002effort,mockus2002two}. So, one of the problems when measuring effort of the overall FOSS development process consists in translating the uneven nature of contributions~\cite{mockus2002two} into a consistent estimation model. Inferring the effort devoted by regular and occasional contributors is not easy, and in addition is a source of inaccuracy. 

The second issue is that the conversion of that data to effort is far from simple, since it is not possible to exactly determine how much time or effort the developer actually spent on those activities. In general, the information we can gather consists of points in time (i.e., timestamps) where specific actions have occurred (e.g., when the commit was performed, when the e-mail was sent, when a comment to a bug notification was submitted), but not how long these actions actually took (e.g., performing the changes in the commit, reading and writing the e-mail, debugging the software to add information on the bug).

\section{Empirical approach}
\label{sec:model}
In this section 
we introduce the concepts and terminology at its base (\ref{subsec:concepts}), and the two types of input (\ref{subsec:workex}) that the model uses to evaluate the developers effort. Using an example scenario, we show how to evaluate the threshold $\theta$ to identify full-time developers (\ref{subsec:definitions}).

In addition, we discuss how to estimate the error in our model (\ref{subsec:determining}), and contribute the novel \textit{goodness} performance measure (\ref{sec:_goodness}) that helps in finding the optimal $\theta$. We conclude the section by evaluating the total estimated effort of the example scenario, as well as the relative estimation error (\ref{sec:_total-error}).

\subsection{Concepts and Terminology}
\label{subsec:concepts}

Our model is build on top of following concepts that are well known in the Software Engineering research literature:

\begin{itemize}

 \item \textbf{Commit:} action by which a developer synchronizes a set of changes to the versioning system repository. A commit is given by a point in time (the timestamp can be obtained from its metadata). 


 \item \textbf{Author (or developer, or contributor):} individual who contributes the changes, but may
not be the one who performs the commit (which is done by the committer). 

 \item \textbf{Committer:} developer who actually performs the commit to the repository, but may have not been the real author of the changes. In our model, we do not consider \emph{committers}, since it would decrease its accuracy.


 \item \textbf{Full-time developer:} Developer who devotes the relative amount of time of a 40 hour-week to the project\footnote{In some countries there is a different work week, e.g., in the UK it is 35 hours. The model presented can be adjusted if needed.}. In some research literature on FOSS, full-time developers are called ``professional'' developers as compared to volunteers~\cite{sowe2008understanding,steinmacher2015social,von2003community}. In this paper, we prefer full-time and non-full-time developer, as the latter can be professionals as well, just with a minor participation in the project.

 \item \textbf{Person-Month:} measure of effort. A person-month is observed when a \emph{full-time developer} devotes 1 month of effort to the project.
\end{itemize}

In addition, for our model we use the following terminology:

\begin{itemize}
  \item \textbf{Period of study n}: timespan (in months) during which the activity (i.e., commits) by developers is aggregated.

  \item \textbf{Threshold $\theta$}: minimum amount of activity (i.e., commits) during the \emph{period of study} by a developer to be considered as a \emph{full-time developer}.
  


  \item \textbf{Effort $E_d$ in a period of study (by developer \textit{d})}: any developer who performs more commits than $\theta$ in a \emph{period of study} of \emph{n} months will be considered as full-time with \emph{n} person-months. A developer who committed \emph{a} commits (with \emph{a} $<$ $\theta$), will be assigned an effort of $\emph{n}*\frac{\emph{a}}{\theta}$ person-months.


  \item \textbf{Effort $E_n$ in a period of study n (for the whole project)}: The overall effort $E_n$ is evaluated by summing up the effort in a period of study of all developers (full-time and non-full-time). The formula to obtain the overall effort is as follows:  
  
  \begin{align*}
    E_n &= \sum_{d=1}^{d_{ft}} n + \sum_{d=1}^{d_{nft}} n*\frac{a_d}{\theta}\\
  \end{align*}
  
  where $d_{ft}$ is the number of full-time developers, $d_{nft}$ the number of non-full-time developers, and $a_d$ the number of commits that a non-full-time developer has contributed in the period $n$, smaller than $\theta$.
  
  \item \textbf{Maximum possible effort M in a period of study (for the whole project)}: This is the effort when all developers are considered full-time, e.g., $\theta=1$. It can be considered as an upper bound of the total effort, and is calculated as follows:

  \begin{align*}
    Max(E_n) &= \sum_{d=1}^{d_{ft} + d_{nft}} n\\
  \end{align*}

\end{itemize}

\subsection{Surveying developers}
\label{subsec:workex}

The first basic input to our model is the self-identification of developers as full- or non-full-time.
This is done by means of a short survey that tries to obtain this information.

One of the questions of the survey explicitly asks respondents to classify themselves in one or the other category (`\textit{What do you consider yourself in the project? (full-time, part-time, occasional contributor}'). 
But as this question can be easily misunderstood, we recommend to use additional, alternative questions that allow to check the input given by developers for consistency.
Thus, we asked them how many hours per week they have usually devoted in the last months to the project (`\textit{On average, how many hours in a week
have you spent in the project in the last six months?}; developers could choose among following options: $>$40h, 40h, 30h, 20h, 10h, $<$5h).

In addition, on a separate web page of the survey, we also showed developers a chart containing their own personal activity (in number of commits per month).
We added a line to the chart with the estimated value of $\theta$, and asked them if we had identified them correctly.
Developers could answer in an open text box. 
Although the aim of this question was to see if developers corroborate our findings, we did not achieve it.
This was because (1) our estimated value of $\theta$ was far away from the value finally obtained; 
and (2) developers were not aware of the details of the model and commented on other issues of lesser interest\footnote{Details can be found in the reproduction package in an anonymized way}.
However, the feedback received clearly indicated that using a period of study of 1 month was not a good choice.
For the analysis presented here, we have set 6 months as the most convenient ``Period of study n'' , as this is the usual time between releases for many projects~\cite{michlmayr2015and}. By doing this, all cycles of six months will be affected by the same process (development, feature freeze, release, etc.).

\subsection{Determining the threshold value $\theta$}
\label{subsec:definitions}

Our effort estimation model depends on how accurately we can identify full-time and non-full-time developers. To illustrate its evaluation, we use an exemplary scenario: let's assume that the development team of a software project is made up of 8 developers, and that they responded to the questionnaire (``\textit{Are you a full-time or a non-full-time developer?}'') as in the second column of Table~\ref{tbl:_example}. In the third column, we report the amounts of commits that each worked on during the period \textbf{n}, as extracted by parsing the versioning system of their project~\cite{robles2004remote}.

\begin{table}[ht!]
\centering
\begin{tabular}{ccc}
       &  &  \\
Dev \# & \vtop{\hbox{\strut Full-time or non-full-time?}\hbox{\strut (\textit{From questionnaires})}}  & \vtop{\hbox{\strut Effort (\# commits)}\hbox{\strut (\textit{From data analysis})}}\\\hline
D1   & F                 & 12     \\
D2   & F                 & 10     \\
D3   & F                 & 13     \\
D4   & NF                & 3      \\
D5   & NF                & 11     \\
D6   & NF                & 8      \\
D7   & F                 & 10     \\
D8   & NF                & 5      \\
\end{tabular}
\caption{Example scenario to determine $\theta$: developers identify themselves as either full-time (F) or non-full-time (NF). We evaluate their effort by counting the number of their commits.}
\label{tbl:_example}
\end{table}

The first three developers (D1, D2 and D3) self-identified as full-time (F). From the data analysis of their commits, we detected that these developers actually worked on 12, 10 and 13 commits respectively during the period under study (for example, a month). Using the data analysis, and if only these three developers were active, the minimum $\theta$ to identify a full-time developer would be 10 commits in the period.

On the other hand, when we consider all the other developers, especially D5, the threshold $\theta=10$ becomes less effective: D5 classifies his/her work as non-full-time, but he/she is responsible for 11 commits in the period \textbf{n}. If we use $\theta=10$ to detect full-time developers also for D5, we wrongly classify him/her as full-time; if, on the contrary, if we use $\theta=11$ to identify full-time developers (and to correctly classify D5), we wrongly classify D2 and D7, who both contributed 10 commits, but self-identified themselves as full-time. Determining the value of $\theta$ becomes therefore an exercise to minimise the classification errors.

\subsection{Determining the estimation error}
\label{subsec:determining}

In the example above, we have the typical information retrieval problem with true/false positives and true/false negatives. In particular, and depending on how the threshold $\theta$ has been chosen:


\begin{itemize}

\item A \textbf{true positive} ($t_p$) 
means that, using a specific $\theta$, the developer identified as full-time is indeed full-time. Using $\theta=10$, the example above comprises 4 true positives (e.g., D1, D2, D3 and D7).

\item A \textbf{false positive} ($f_p$) 
means that the developer is flagged as full-time, but s/he is not. Using $\theta=10$, the example above comprises 1 false positive (e.g., D5).

\item A \textbf{false negative} ($f_n$) 
means that the developer has not been identified as full-time, albeit s/he is. Using $\theta=10$, the example above does not comprise any false negatives. If we used $\theta=11$, instead of $\theta=10$, the example above would contain 2 false negatives (e.g., D2 and D7), but no false positives.

\item A \textbf{true negative} ($t_n$) 
means that a non-full-time contributors has been correctly flagged. Using $\theta=10$, the example above contains 3 true negatives (e.g, D4, D6 and D8).

\end{itemize}

\begin{figure*}[ht!]
\begin{center}
\includegraphics[keepaspectratio=true,width=0.79\textwidth]{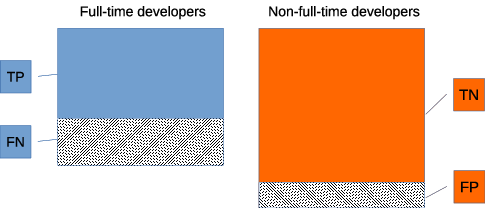}
\caption{Visual representation of the tp, fn, fp and tn values for the full-time and non-full-time developer sets}
\label{fig:_tp_fn_fp_tn}
\end{center}
\end{figure*}

The visual representation of these terms is displayed in Figure~\ref{fig:_tp_fn_fp_tn}, using the full-time and non-full-time subsets of developers. Within the realm of information retrieval, we can also evaluate the values of well-known performance measures:
\begin{itemize}

\item The \textbf{precision} of the model is the ratio of true positives (i.e., the correctly identified full-time developers) and the number of all the identified (correctly or not) full-time developers, or $P_d = \frac{t_p}{t_p + f_p}$.

\item The \textbf{recall} of the model (or its sensitivity) is the ratio of correctly identified full-time developers and the total number of full-time developers (whether they have been correctly identified or not), as in $R_d = \frac{t_p}{t_p + f_n}$.

\item The \textbf{accuracy} is the ratio of all the correctly predicted developers, and the overall set of observations: $A_d = \frac{t_p + t_n}{t_p + t_n + f_p + f_n}$.

\item The \textbf{F-measure} is a combination of both precision and recall, $F_d = 2 * \frac{P_d * R_d}{P_d + R_d}$.

\end{itemize}

However, these `classic' performance measurements do not take into account the fact that classification errors can compensate each other: depending on the project, the same $\theta$ could classify a non-full-time developer as full-time (e.g., a false positive) and at the same time a full-time developer as non-full-time (e.g., a false negative). In this case, we state that the two errors \textit{compensate} each other.

\subsection{Compensating the classification errors: the \textit{goodness} measure}
\label{sec:_goodness}

The formulation of the \emph{goodness} measurement is designed to explicitly consider the compensation of classification errors: this fact alone sets our model apart from traditional information retrieval techniques. Below we illustrate (i) how the \textit{goodness} performance measure can accommodate the compensation of classification errors, and (ii) how it is pivotal in selecting the most probable threshold $\theta$ to detect the full-time developers for our model.



The evaluation of the \textit{goodness} measurement stems from the intrinsic distribution of the developers' work patterns. Depending on their contributions and what type of contributors they consider themselves to be, a larger or smaller number of wrongly classified developers could seriously skew the evaluation of the overall effort. It is also worth noticing that there is currently no mention in the literature of the possibility of classification errors, and the need of error compensation. As it stands, this is the first research work that addresses this open issue. The formula of the \textit{goodness} measure is as follows:

\begin{itemize}

\item \textbf{Goodness} = $1-\frac{|(t_p + f_p)- (t_p + f_n)|}{t_p + f_n + f_p}$ = $1-\frac{|f_p - f_n|}{t_p + f_n + f_p}$

\end{itemize}

The \textit{goodness} measure depends on the number of \textit{non-compensated} classifications (the numerator, an absolute value, since compensation works both ways), divided by the total number of positives and false negatives. Subtracting this fraction to 1, the \emph{goodness} becomes larger as long as false positives and negatives have been compensated (or if no compensation was necessary). The measure of goodness could also consider the \emph{true negatives} ($t_n$) instead of the \emph{true positives} ($t_p$). The reason of not including ($t_n$) is due to the typical distribution of FOSS developers: a large number of FOSS developers have an extremely low level of contribution. As a result, including a large number of $t_n$ into the definition of goodness would inevitably lower its discriminative power, as compared to using $t_p$.

Using the same example scenario depicted in Table~\ref{tbl:_example}, we report in Table~\ref{tbl:_example_goodness} the values of $t_p$, $f_p$, $f_n$ and \textit{goodness}, using different values of $\theta$ (that is, number of commits). As visible, the relatively high number of false positives ($f_p$) has a negative impact on the goodness measure, especially when $\theta$ is small. As long as more false positives get compensated by false negatives, the goodness measure finds its maximum: we posit that the best values of $\theta$ lie where the goodness is maximised (highlighted in Table~\ref{tbl:_example_goodness}).

\begin{table}[!ht]
\centering
\begin{tabular}{r|cccc}
$\theta$ & $t_p$ & $f_p$  & $f_n$ & \textit{goodness} \\\hline
1  & 4 & 4 & 0 & 0.50 \\
2  & 4 & 4 & 0 & 0.50 \\
3  & 4 & 4 & 0 & 0.50 \\
4  & 4 & 3 & 0 & 0.57 \\
5  & 4 & 3 & 0 & 0.57 \\
6  & 4 & 2 & 0 & 0.67 \\
7  & 4 & 2 & 0 & 0.67 \\
8  & 4 & 2 & 0 & 0.67 \\
\rowcolor{LightCyan} 9  & 4 & 1 & 0 & 0.80 \\
\rowcolor{LightCyan} 10 & 4 & 1 & 0 & 0.80 \\
\rowcolor{LightCyan} 11 & 2 & 1 & 2 & 0.80 \\
12 & 2 & 0 & 2 & 0.50 \\
13 & 1 & 0 & 3 & 0.25
\end{tabular}
\caption{True positives, false positives and false negatives (as well as the resulting goodness measure) at different values of $\theta$. The data is based on the example from Table~\ref{tbl:_example}.}
\label{tbl:_example_goodness}
\end{table}

\subsection{Total estimated effort and estimation error}
\label{sec:_total-error}
The identification of $\theta$ allows to evaluate the effort by developers for a certain period, or for the entire project. In the latter case, an adjustment of the threshold might become necessary. In the example scenario that was discussed above (see Table~\ref{tbl:_example} and~\ref{tbl:_example_goodness}), the threshold that maximises the \textit{goodness} measure lies between 9 and 11. If we choose $\theta=10$, the overall effort by developers is 6.6 PM, as in the sum of:
\begin{itemize}
    \item 5 PM (1 PM each for the D1, D2, D3, D5 and D7 developers)\footnote{Full-time developers are assumed to work 1 PM every month, even if they work more than the chosen $\theta$. We call this aspect the \textit{saturation} of a developer's effort.}
    \item $\frac{3}{10}$ PM (for the D4 developer)
    \item $\frac{8}{10}$ PM (for the D6 developer)
    \item $\frac{5}{10}$ PM (for the D8 developer)
\end{itemize}

Choosing a different threshold, but still within the values that maximise the \textit{goodness} function (i.e., the shaded rows in Table~\ref{tbl:_example_goodness}), has a small effect on the overall effort, that sums up to 6.78 PM in case of $\theta=9$, and 6.27 PM in case of $\theta=11$. The estimation errors grow larger as long as one moves away from the `safe' range of $\theta$. The sum to evaluate the overall effort, as presented above for the example scenario, is the same that was used to evaluate the effort of the projects under study.

Table~\ref{tbl:_example_error} lists the estimation errors (in percentage) if a different $\theta$ is selected, as compared to the $\theta=10$ that maximises the \textit{goodness} of our example scenario.

\begin{table}[!ht]
\centering
\begin{tabular}{r|cr}
$\theta$  & Effort (PM)   & Estimation error \\\hline
1  & 8.00    & 21.21\%          \\
2  & 8.00    & 21.21\%          \\
3  & 8.00    & 21.21\%          \\
4  & 7.75 & 17.42\%          \\
5  & 7.60 & 15.15\%          \\
6  & 7.33 & 11.11\%          \\
7  & 7.14 & 8.23\%           \\
8  & 7.00    & 6.06\%           \\
\rowcolor{LightCyan} 
9  & 6.78 & 2.69\%           \\
\rowcolor{LightCyan} 
10 & 6.60 & --           \\
\rowcolor{LightCyan} 
11 & 6.27 & -4.96\%          \\
12 & 5.92 & -10.35\%         \\
13 & 5.54 & -16.08\%        
\end{tabular}
\caption{Values of global effort (in PM) by developers, and the percentages of error when choosing a different threshold. The data is based on the example from Table~\ref{tbl:_example}.}
\label{tbl:_example_error}
\end{table}

\section{Deployment of the model}
\label{sec:projects}

The model presented above is deployed using 6 large FOSS projects that have a different degree of participation of commercial companies. The selection of these projects is based on convenience sampling: two of the authors of this paper have a direct contact with the communities under study.

All selected projects follow common practices found nowadays in large FOSS software development projects: they have code review in process (using an external tool like Gerrit\footnote{\url{https://www.gerritcodereview.com/}}, except Ceph that uses GitHub's pull-request driven process) and recommend to do commit squashing~\cite{kalliamvakou2014promises}. 
Squashing a commit refers to move changes introduced in a commit into its parent, to end up with one commit instead of two.
If you repeat this process multiple times, n commits can be reduced (i.e., `squashed') to a single one. 
It is common practice in many FOSS projects to squash all commits that belong to a change, after it is reviewed and accepted~\cite{kononenko2018studying}.

\begin{table}[!th]
\caption{Main parameters of the studied projects, January 2014.}
\center
\begin{tabular}{|l|c|c|p{1.5cm}|p{1.5cm}|c|}
 \hline
 Project & Type & First release & \# Authors \newline (distinct) & \# Commits \newline (w/o bots) & SLOC \\ \hline
 Ceph & Consortium & 07/2008 (v0.2) & \multicolumn{1}{r|}{158}  & \multicolumn{1}{r|}{37,254} & $418K$ \\ \hline
 Linux & Consortium & 08/1991 & \multicolumn{1}{r|}{3,665} & \multicolumn{1}{r|}{540,263} & $>10M$ \\  \hline
 Moodle & Organization & 08/2002 (v1.0) & \multicolumn{1}{r|}{174}  & \multicolumn{1}{r|}{52,523} & $1.28M$ \\ \hline
 OpenStack & Consortium & 10/2010 &  \multicolumn{1}{r|}{1,410} & \multicolumn{1}{r|}{68,587}  & $1.65M$ \\ \hline
 WebKit & Consortium & 06/2005 & \multicolumn{1}{r|}{690} & \multicolumn{1}{r|}{107,471} & $5.0M$ \\ \hline
 MediaWiki & Organization & 01/2001 & \multicolumn{1}{r|}{605} & \multicolumn{1}{r|}{357,786} & $340K$ \\ \hline
\end{tabular}
\label{tab:general}
\end{table}

Summary statistics are provided in Table~\ref{tab:general}, with an additional column stating whether it is a consortium with commercial participation (`consortium'), or an organization (commercial or not) driving the project (`organization'). 

\paragraph{OpenStack} -- OpenStack is a software project to build a SaaS (software as a service) platform. Over 200 companies have become involved in the project. AMD, Brocade Communications Systems, Canonical, Cisco, Dell, Ericsson, Groupe Bull, HP, IBM, InkTank, Intel, NEC, Rackspace Hosting, Red Hat, SUSE Linux, VMware Yahoo! and many others have been participating in OpenStack.

\paragraph{Moodle} -- Moodle is an FOSS project that implements a PHP-based learning management system: it started as a volunteer project in 2002, but it has since evolved into a company-driven FOSS project. Moodle HQ is the company that keeps developing it, and there is an estimated 80 international companies that help in its development.

\paragraph{Ceph} -- Ceph is a freely available storage platform\footnote{\url{https://github.com/ceph/ceph}}, and it also started as a volunteer based FOSS project. After being merged into the Linux kernel in 2010, it was incorporated by RedHat that is now responsible for its in-house development. Various large organisations contribute to its development, from Canonical to Cisco, Fujitsu and Intel.

\paragraph{Linux} -- Linux represents the most well-known success story for FOSS projects. Started as a volunteer project back in 1991, Linux grew exponentially, although it is still a community-driven project. The amount of contributions from commercial companies have slowly outgrown those of the individual developers: the latter account now for only 10\% of the developers of the operating system.

\paragraph{WebKit} -- WebKit is an FOSS project started by Apple, although it started as a fork of an existing KDE project. It implements the engine that sits behind the Safari web browser. KDE, as well as other FOSS communities, still participates in the development of WebKit, but large commercial organisations (Google, Nokia) have also contributed to it. In April 2013, Google forked the WebKit creating to Blink. Since then it is mainly a project backed by Apple.

\paragraph{MediaWiki} -- Wikimedia is the non-profit organisation that supports the growth of the Wikipedia online encyclopedia. Wikimedia is directly in charge of the MediaWiki FOSS project\footnote{\url{https://phabricator.wikimedia.org/source/mediawiki/}}, that implements the wiki engine that allows pages to be added, modified and contributed to by anyone.


\section{Data Gathering}
\label{sec:experiment}

As mentioned above, the estimation model that we propose in this paper obtains data from two sources: (i) the VCS of the project under study, and (ii) the questionnaire data that was sent to the developers identified in the VCS of each project. Below we describe the steps that were performed to gather the data from each source.

\subsection{Versioning System Data}
\label{sec:_dev-data}

The information from the VCS is obtained by means of Perceval~\cite{duenas2018perceval}, which is part of the MetricsGrimoire toolset\footnote{\url{http://metricsgrimoire.github.io}}. Perceval retrieves development information and metadata from repositories\footnote{The Ceph, Linux, Moodle, MediaWiki and OpenStack datasets were retrieved from git repositories. The WebKit dataset was retrieved from a Subversion repository.} of the project, and provides it as JSON files, ready for analysis.

The data obtained from the VCS repositories was subject to two cleaning operations: one related to the extraction and reconciliation of \textit{duplicated} developer identities~\cite{robles2005developer,wiese2016mailing}; and one related to the exclusion of commits generated by \textit{automated bots}. We performed those operations for all the projects in the sample.

Using OpenStack as an example, from the analysis of its commits we observed that there are 1,840 distinct author identifiers in the VCS of OpenStack. After applying a merging algorithm~\cite{kouters2012s}, manual inspection and feedback from the community, 1,410 distinct real authors have been identified, once duplicate IDs were merged. Additionally, the total number of commits is almost 90,000, although many of these have been done automatically by bots. The committers identified as bots, as well as their commits, were excluded from the analysis.
Bots are responsible for 21,284 commits, almost 25\% of all commits in the project: those bots and their commits were excluded from the data used to evaluate $\theta$.



OpenStack, as many other FOSS projects, has an uneven contribution pattern where a small group of developers have authored a major part of the project. In summary, 80\% of the commits have been authored by slightly less than 8\% of the authors, while 90\% of the commits correspond to about 17\% of all the authors. In addition, as the corporate involvement in OpenStack is significant, this should allow us to identify full-time developers from companies: the OpenStack Foundation estimates that around 250 developers work professionally in the project.

At the end of this data analysis, for each project in our sample we obtained the number of monthly commits per developer.


\subsection{Online survey of developers effort}

To obtain data of the time commitment of the FOSS developers in the sampled projects, we designed an on-line survey (as described in Section~\ref{subsec:workex}).
We obtained the e-mail addresses of the developers from their authorship information in the VCS, and sent them an e-mail with an invitation to participate.

Gathering all the distinct developers from the 6 sampled FOSS projects, an overall 6,700 personalized mails were sent, and the survey was answered by a grand total of 1,028 respondents. We removed 55 survey responses because they were empty or we could see from the answer in the free text box that the respondent had misunderstood/misinterpreted the questions (e.g., there was a developer referring to his contributions to the LibreOffice project, which is not part of the analysed sample of projects). We also amended 32 surveys (e.g., some respondents stated to be professional developers hired by a company and devoted 40 or more hours a week to the project, but had left empty their status, which we set to ``full time'').

Table~\ref{tab:emails_sent} reports the number of responses (and the ratio) obtained from the developers of the sampled projects.

\begin{table}[!ht]
\caption{Summary of emails sent to the developers of the sampled projects, and ratio of completed questionnaires}
\center
\begin{tabular}{l|rrr}
          & Sent  & Received & Ratio   \\ \hline
Ceph      & 158   & 24       & 15.19\% \\ \hline
Linux     & 3,665 & 652      & 17.79\% \\ \hline
Moodle    & 174   & 42       & 24.14\% \\ \hline
OpenStack & 1,407 & 131      & 9.31\%  \\ \hline
WebKit    & 690   & 85       & 12.32\% \\ \hline
MediaWiki & 605   & 94       & 15.54\% \\ \hline\hline
\rowcolor{LightCyan} 
Total     & 6,699 & 1,028    & 15.35\% \\ \hline

\end{tabular}
\label{tab:emails_sent}
\end{table}

\section{Model deployment -- OpenStack}
\label{sec:results}
This section reports the results of our analysis: we first present an in-depth analysis of the OpenStack project, and later an overall view of all the projects in the sample.

%
%
%

\subsection{Identification of full-time developers}
\label{sec:identification}

In this section we describe how the full-time developers were identified within the OpenStack contributors, by using the threshold $\theta$ and calibrating the results of the questionnaire. The same approach was followed for the other projects in the sample.

As a first step we analysed the questionnaires: we gathered that 37 developers identify themselves as full-time, while 54 developers consider themselves as non-full-time. As a second step, we extracted the number of commits of each developer, in the six months prior the questionnaire. Given those two sources of information, we need to answer the question: what is the \textit{minimum} number of commits that more precisely separates full-time from non-full-time developers?

\begin{figure}[ht!]
\begin{center}
\includegraphics[keepaspectratio=true,width=0.89\textwidth]{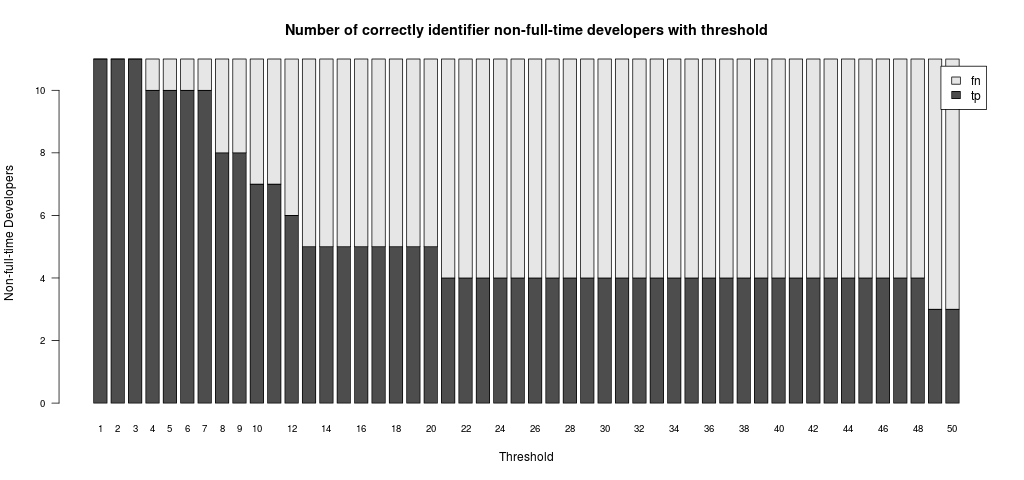}
\includegraphics[keepaspectratio=true,width=0.89\textwidth]{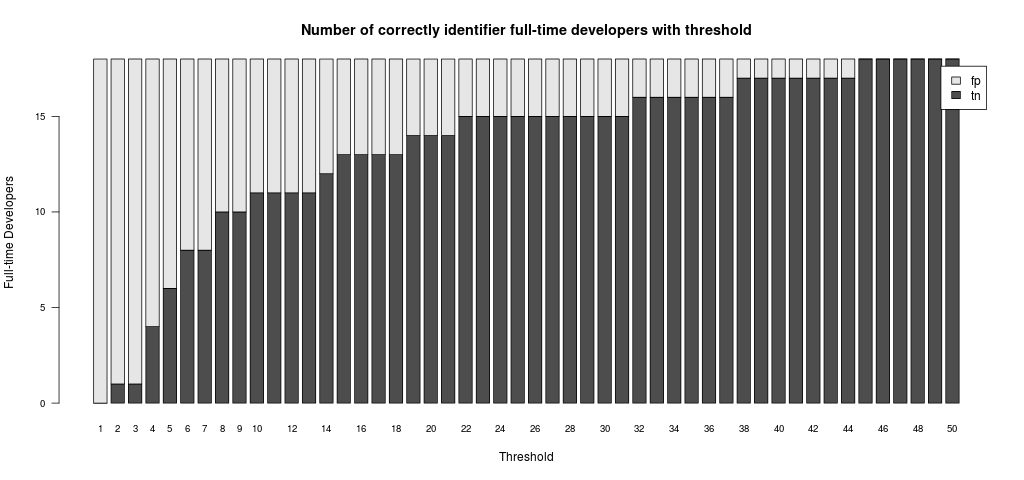}

	\caption{(Top) True positives (tp) and false negatives (fn) for full-time developer identification for several values of $\theta$ (OpenStack project). (Bottom) False positives (fp) and true negatives (tn) for non-full-time developer identification for several values of $\theta$. The dimension of the vertical axis has been maintained to ease the comparison of the size of both populations.}
	\label{fig:_tp_fn_tn_fp_openstack}
\end{center}
\end{figure}

To answer that question, the two plots in Figure~\ref{fig:_tp_fn_tn_fp_openstack} classify the developers as following (at a certain level of $\theta$):
\begin{itemize}
    \item \textit{true positives}: number of full-time OpenStack developers that are classified as full-timers; 
    \item \textit{false negatives}: number of full-time OpenStack developers wrongly classified as non-full-timers; 
    \item \textit{false positives}: number of non-full-time OpenStack developers that were classified as full-timers; 
    \item \textit{true negatives}: number of non-full-time OpenStack developers correctly classified as non-full-timers.
\end{itemize}

\paragraph{Full-time developers --} When $\theta$ is set to 1, developers who contribute at least one commit in a six-month time-span will be considered as full-time by our algorithm. From Figure~\ref{fig:_tp_fn_tn_fp_openstack}, the precision to detect a full-time developer with $\theta=1$ is maximum: this means that each of the 37 self-identified full-time developer has contributed at least one commit during the allocated time-span (i.e., six months). 

The number of true positives (tp) decreases once $\theta$ gets larger: for example, out of the 37 self-identified full-timers, only 30 developers contribute at least 8 commits every six months (i.e., $\theta=8$). The remaining 7 developers are classified as false negatives: even if they are full-time developers, our approach considers them as non-full-timers. In this case, the precision of using a threshold $\theta=8$ to detect full-time developers is $\frac{30}{37}$.

\paragraph{Non-full-time developers --} The same approach applies to non-full-time developers, but based on false positives and true negatives (see bottom plot of Figure~\ref{fig:_tp_fn_tn_fp_openstack}): setting $\theta=1$, all the self-identified non-full-time developers are false positives, and our algorithm considers them as full-timers. 

As seen in the bottom plot of the same Figure, also the number of true negatives (tn) increases, as long as $\theta$ gets larger. At $\theta=8$, for example, 38 out of 54 non-full-time developers are correctly identified as non-full-timers (i.e., true negatives); while 6 out of 54 non-full-time developers are still considered as full-timers, at that level of $\theta$ (i.e., false positives).

\subsection{Compensation of error and the goodness metric}
\label{sec:compensation}

\begin{figure}[ht!]
\begin{center}
\includegraphics[keepaspectratio=true,width=0.80\textwidth]{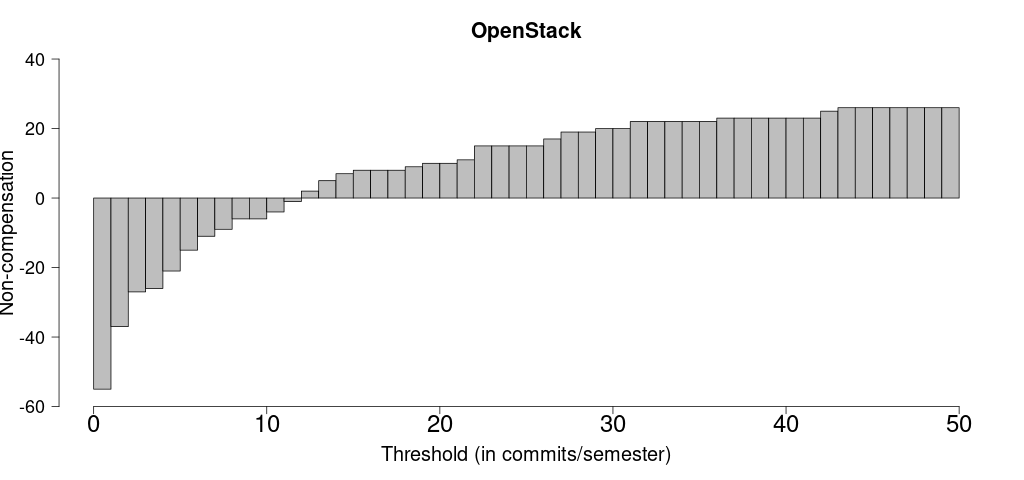}
	\caption{Compensation between false positives and false negatives for values of t. Negatives values of the y axis indicate false positives that do not get compensated by false negatives, while positive values of the y axis indicate false negatives not compensated by false positives.}
	\label{fig:_compensation_OpenStack}
\end{center}
\end{figure}

Figure~\ref{fig:_compensation_OpenStack} displays graphically the compensation pattern in the OpenStack project, and for different values of $\theta$. The number of wrongly classified non-full-time developers are subtracted from the number of wrongly classified full-time developers. 

As it can be seen, for $\theta=1$ no full-time developer has been incorrectly classified, but all non-full-time developers have. No compensation occurs then. However, as $\theta$ increases, the error of incorrectly classifying full-time developers is compensated by the error of incorrectly classifying non-full-time developers. The value for which the number of non-compensated errors is minimum corresponds to t=12, where the difference is 1 (i.e., there is one false negative more than false positives).





\begin{figure}[ht!]
\begin{center}
\includegraphics[keepaspectratio=true,width=0.98\textwidth]{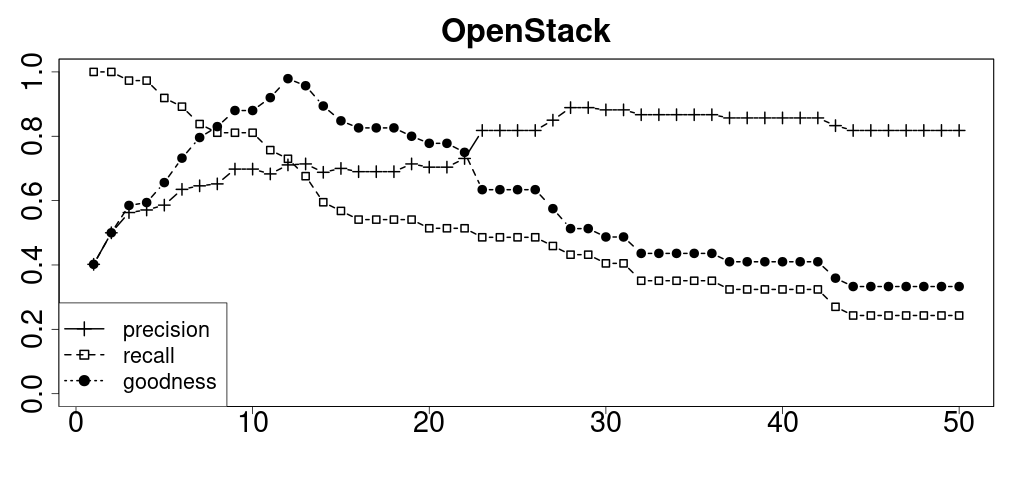}
\caption{Relevance of results (for many threshold values) for the OpenStack project.}
\label{fig:_relevance_OpenStack}
\end{center}
\end{figure}

Figure~\ref{fig:_relevance_OpenStack} shows that, for the OpenStack project, the value of goodness peaks at $\theta=12$ (0.979). Thus, the error for our effort estimation should be smaller for $\theta=12$, due to the effect of compensation that the goodness factor recognizes.

\label{sec:threshold}

\subsection{Effort estimation}
\label{sec:estimation}

Having obtained feedback from the OpenStack developers through the survey, the identification of full-time developers via $\theta$ becomes more precise. Depending on the $\theta$ chosen, the effort estimation will provide more or less accurate values.

Table~\ref{tab:openstack-effort} shows the results of the estimated effort in OpenStack, and using several values of $\theta$. Considering the global effort through the whole lifespan of the project, we can see that -as expected- the number of person-months decreases while the threshold increases: less and less developers are considered as full-time, thus pushing the effort down. The upper bound of effort for OpenStack, where a single commit in a semester would have been coded as 6 person-months, results in 17,400 person-months. 

\begin{table}[!h]
\center
\caption{Estimated total effort (in PM) for the OpenStack project, given per threshold value (column $\theta$). Estimated effort values for all semesters. The darker the background, the higher the value of goodness ($white < 0.8 \leq light blue < 0.9 \leq greenblue \leq 1.0$)}
\begin{tabular}{r|r|cccccccc}
& & \multicolumn{7}{c}{Effort (semester, PM)}\\
$\theta$ & Effort (total, PM) & 10s2 & 11s1 & 11s2 & 12s1 & 12s2 & 13s1 & 13s2 \\ \hline
\rowcolor{goldenyellow}1 & 17,400 & 426 & 816 & 1,242 & 1,890 & 2,742 & 4,368 & 5,916 \\ 
2 & 15,255 & 420 & 813 & 1,170 & 1,647 & 2,385 & 3,783 & 5,037 \\ 
3 & 13,804 & 402 & 798 & 1,110 & 1,486 & 2,140 & 3,384 & 4,484 \\ 
4 & 12,747 & 389 & 785 & 1,056 & 1,383 & 1,964 & 3,074 & 4,098 \\ 
5 & 11,891 & 378 & 768 & 1,008 & 1,295 & 1,830 & 2,821 & 3,791 \\ 
6 & 11,213 & 370 & 757 & 964 & 1,224 & 1,726 & 2,618 & 3,554 \\ 
7 & 10,629 & 362 & 743 & 927 & 1,164 & 1,635 & 2,451 & 3,346 \\
\rowcolor{0.83} 8 & 10,128 & 355 & 732 & 896 & 1,112 & 1,560 & 2,310 & 3,164 \\ 
\rowcolor{0.88} 9 & 9,683 & 345 & 721 & 866 & 1,065 & 1,493 & 2,189 & 3,003 \\ 
\rowcolor{0.88} 10 & 9,298 & 337 & 711 & 841 & 1,025 & 1,434 & 2,086 & 2,864 \\ 
\rowcolor{0.92} 11 & 8,957 & 330 & 700 & 819 & 987 & 1,381 & 1,997 & 2,743 \\ \hline
\rowcolor{0.98} 12 & 8,655 & 324 & 690 & 800 & 955 & 1,334 & 1,919 & 2,634 \\ \hline
\rowcolor{0.96} 13 & 8,370 & 318 & 680 & 782 & 924 & 1,289 & 1,847 & 2,531 \\ 
\rowcolor{0.89} 14 & 8,112 & 313 & 672 & 767 & 896 & 1,247 & 1,781 & 2,437 \\ 
\rowcolor{0.85} 15 & 7,876 & 308 & 663 & 753 & 872 & 1,208 & 1,721 & 2,351 \\ 
\rowcolor{0.83} 16 & 7,662 & 303 & 656 & 740 & 850 & 1,173 & 1,666 & 2,275 \\ 
\rowcolor{0.83} 17 & 7,466 & 298 & 648 & 729 & 830 & 1,140 & 1,615 & 2,206 \\ 
\rowcolor{0.83} 18 & 7,284 & 294 & 641 & 719 & 812 & 1,107 & 1,568 & 2,142 \\ 
\rowcolor{0.80} 19 & 7,109 & 291 & 634 & 708 & 794 & 1,077 & 1,522 & 2,083 \\ 
20 & 6,945 & 288 & 628 & 698 & 777 & 1,048 & 1,481 & 2,025 \\ 
\end{tabular}
\label{tab:openstack-effort}
\end{table}

According to the analysis, the best values for $\theta$ are between 9 and 12. The highlighted value of $\theta=12$ in Table~\ref{tab:openstack-effort} is what minimizes the error of considering developers as full-time using the feedback from the OpenStack developers. For the $\theta = 12$ scenario, the estimation of the amount of effort that was invested in the OpenStack project lies around 9,000 person-months (750 person-years). This can be considered as the most accurate, and it reaches only half of the estimation value produced in the upper bound estimation (with $\theta=1$).

The table also provides information for every semester. As can be observed, the effort devoted to the OpenStack project is increasing with time, with a maximum for all the thresholds in the second semester of 2013. If we take the value of 12 as the threshold, 2,634 person-months have been worked during the last six months. This implies that we estimate that the actual effort on OpenStack has been around 440 person-months during the last six months. When asked about these figures, OpenStack members confirmed that they sound reasonable as the estimation of the OpenStack foundation is that currently over 250 professional developers work on the project hired by companies\footnote{This type of informal feedback was gathered only from OpenStack developers, but not repeated for the other projects in the sample.}.

On a side note, it should be noted that the ratio between the effort estimated as the upper bound ($\theta=1$) and the one with $\theta=12$ is steadily increasing. In the second semester of 2010 the ratio was 1.31, while for the second semester of 2013 it has grown to 2.25. This is because the number of non-full-time contributors has grown as well. Thus we can see how much the ``minor contributors'' in the project affect the estimations given by our model. While the error (noise) introduced by these small contributions is included using the upper bound, higher values of $\theta$ filter it out. The result is not only a more realistic estimation, but also an estimation where the more error-prone assumptions (i.e., non full-time developers) have a smaller weight in the total effort.

\section{Further model deployment}
\label{sec:_results_replicated}

The estimation model deployed in the OpenStack project was also evaluated for the other systems composing our sample. In this section, we perform the model deployment for 5 other systems to
\begin{enumerate}[label=(\roman*)]
  \item identify their full-time and non-full-time developers (\ref{sec:_identification-further}),
  \item extract the values of $\theta$ where compensation happens more frequently for each system (\ref{sec:_compensation-further}), and
  \item find the goodness value that maximises the accuracy of the (project-specific) model (\ref{sec:_goodness-further}).
\end{enumerate}

\begin{table}[t]
\scriptsize	
\center

\caption{Estimated total effort (in PM) for all projects from 2010s2-2013s2 (3.5 years), given per threshold value (column $\theta$). The darker the background, the higher the value of goodness ($white < 0.8 \leq light blue < 0.9 \leq greenblue \leq 1.0$)}
\begin{tabular}{|r|r||r|r||r|r||r|r||r|r||r|r|}
\hline
$\theta$ & OStack & t & Moodle & t & WebKit & t & Linux & t & Ceph & t & MWiki \\ \hline
\rowcolor{goldenyellow}1 & 17,400 &  1 & 2,688 & 1 &  11,976  & 1 & 57,390 & 1 & 1,056 &  1 & 5,508 \\
... & ... & ... & ... & ... & ... & ... & ... & ... & ... & ... & ... \\
6 & & 8 & & 11 & & \cellcolor{0.81} 11 & \cellcolor{0.81} 32,156 & 18  & 574 & \cellcolor{0.89} 30 & \cellcolor{0.89} 3,348 \\ 
7 & & 9 & & 12 & & \cellcolor{0.85} 12 & \cellcolor{0.85} 31,157 & 19 & 567 & \cellcolor{0.89} 31 & \cellcolor{0.89} 3,324 \\ 
\cellcolor{0.83} 8 & \cellcolor{0.83} 10,128 & 10 & & \cellcolor{0.83} 13 & \cellcolor{0.83} 8,022 & \cellcolor{0.87} 13 & \cellcolor{0.87} 30,240 & 20 & 560 & \cellcolor{0.92} 32 & \cellcolor{0.92} 3,302 \\  
\cellcolor{0.88} 9 & \cellcolor{0.88} 9,683 & \cellcolor{0.83} 11 & \cellcolor{0.83} 1,846 & \cellcolor{0.83} 14 &  \cellcolor{0.83} 7,887  & \cellcolor{0.89} 14 & \cellcolor{0.89} 29,393  & 21 & 554 & \cellcolor{0.92} 33 & \cellcolor{0.92} 3,279\\ 
\cellcolor{0.88} 10 & \cellcolor{0.88} 9,298 & \cellcolor{0.89} 12 & \cellcolor{0.89} 1,803 & \cellcolor{0.87} 15 &  \cellcolor{0.87} 7,753 & \cellcolor{0.92} 15 & \cellcolor{0.92} 28,602  & 22 & 548 & \cellcolor{0.92} 34 & \cellcolor{0.92} 3,258\\ 
\cellcolor{0.92} 11 & \cellcolor{0.92} 8,957 & \cellcolor{0.94} 13 & \cellcolor{0.94} 1,765 & \cellcolor{0.95} 16 & \cellcolor{0.95} 7,619 & \cellcolor{0.96} 16 & \cellcolor{0.96} 27,865  & 23 & 542 & \cellcolor{0.96} 35 & \cellcolor{0.96} 3,237\\ \hline
\cellcolor{0.98} 12 & \cellcolor{0.98} 8,655 & \cellcolor{1.00} 14 & \cellcolor{1.00} 1,731 & \cellcolor{1.00} 17 & \cellcolor{1.00} 7,493 & \cellcolor{0.97} 17 & \cellcolor{0.97} 27,173 & \cellcolor{0.83} 24 & \cellcolor{0.83} 536 & \cellcolor{1.00} 36 & \cellcolor{1.00} 3,217 \\ \hline
\cellcolor{0.96} 13 & \cellcolor{0.96} 8,370 & \cellcolor{0.94} 15 & \cellcolor{0.94} 1,698 & \cellcolor{0.91} 18 & \cellcolor{0.91} 7,372 & \cellcolor{0.97} 18 & \cellcolor{0.97} 26,524 &\cellcolor{0.83} 25 & \cellcolor{0.83} 530 & \cellcolor{1.00} 37 & \cellcolor{1.00} 3,198 \\ 
\cellcolor{0.89} 14 & \cellcolor{0.89} 8,112 & \cellcolor{0.94} 16 & \cellcolor{0.94} 1,666 & \cellcolor{0.91} 19 & \cellcolor{0.91} 7,257 & \cellcolor{0.97} 19 & \cellcolor{0.97} 25,914 & \cellcolor{0.83} 26 & \cellcolor{0.83} 525 & \cellcolor{0.96} 38 & \cellcolor{0.96} 3,179 \\ 
\cellcolor{0.85} 15 & \cellcolor{0.85} 7,876 & \cellcolor{0.94} 17 & \cellcolor{0.94} 1,636 & \cellcolor{0.91} 20 & \cellcolor{0.91} 7,146 & \cellcolor{0.95} 20 & \cellcolor{0.95} 25,334 & \cellcolor{0.83} 27 & \cellcolor{0.83} 520 & \cellcolor{0.96} 39 & \cellcolor{0.96} 3,161 \\ 
\cellcolor{0.83} 16 & \cellcolor{0.83} 7,662 & \cellcolor{0.94} 18 & \cellcolor{0.94} 1,609 & \cellcolor{0.91} 21 & \cellcolor{0.91} 7,037 & \cellcolor{0.93} 21 & \cellcolor{0.93} 24,781 & \cellcolor{0.83} 28 & \cellcolor{0.83} 516 & \cellcolor{0.96} 40 & \cellcolor{0.96} 3,142 \\  
\cellcolor{0.83} 17 & \cellcolor{0.83} 7,466 & \cellcolor{0.87} 19 & \cellcolor{0.87} 1,584 & \cellcolor{0.91} 22 & \cellcolor{0.91} 6,933 & \cellcolor{0.92} 22 & \cellcolor{0.92} 24,256 & \cellcolor{0.83} 29 & \cellcolor{0.83} 511  & \cellcolor{0.91} 41 & \cellcolor{0.91} 3,124 \\  
\cellcolor{0.83} 18 & \cellcolor{0.83} 7,284 & \cellcolor{0.87} 20 & \cellcolor{0.87} 1,561 & \cellcolor{0.91} 23 & \cellcolor{0.91} 6,833 & \cellcolor{0.90} 23 & \cellcolor{0.90} 23,761 & \cellcolor{0.83} 30 & \cellcolor{0.83} 506 & \cellcolor{0.91} 42 & \cellcolor{0.91} 3,106 \\
19 & & 21 & & \cellcolor{0.91} 24 & \cellcolor{0.91} 6,737 & \cellcolor{0.87} 24 & \cellcolor{0.87} 23,290 & \cellcolor{0.83} 31 & \cellcolor{0.83} 501 & \cellcolor{0.86} 43 & \cellcolor{0.86} 3,089 \\ 
20 & & 22 & & \cellcolor{0.91} 25 & \cellcolor{0.91} 6,646 & \cellcolor{0.85} 25 & \cellcolor{0.85} 22,839 & \cellcolor{0.83} 32 & \cellcolor{0.83} 496 & \cellcolor{0.86} 44 & \cellcolor{0.86} 3,072 \\ 
21 & & 23 & & \cellcolor{0.85} 26 &  \cellcolor{0.85} 6,558 & \cellcolor{0.84} 26 & \cellcolor{0.84} 22,411 & \cellcolor{0.83} 33 & \cellcolor{0.83} 492 & \cellcolor{0.86} 45 & \cellcolor{0.86} 3,056 \\ 
22 & & 24 & & \cellcolor{0.85} 27 & \cellcolor{0.85} 6,472 & \cellcolor{0.82} 27 & \cellcolor{0.82} 22,002 & \cellcolor{0.83} 34 & \cellcolor{0.83} 488 & \cellcolor{0.86} 46 & \cellcolor{0.86} 3,040 \\ 
23 & & 25 & & \cellcolor{0.85} 28 & \cellcolor{0.85} 6,390 & \cellcolor{0.82} 28 & \cellcolor{0.82} 21,609 & \cellcolor{0.83} 35 & \cellcolor{0.83} 484 & \cellcolor{0.86} 47 & \cellcolor{0.86} 3,025 \\  
\hline
\end{tabular}
\label{tab:all-effort}
\end{table}

Table~\ref{tab:all-effort} shows the summary of the calculations of both the $\theta$ value, and the corresponding effort for the other projects in the sample. In the next sections we describe how the estimation model was deployed in each of the additional systems in the sample.

\subsection{Identification of full-time developers}
\label{sec:_identification-further}

\begin{figure*}[ht!]
\begin{center}
\includegraphics[keepaspectratio=true,width=0.49\textwidth]{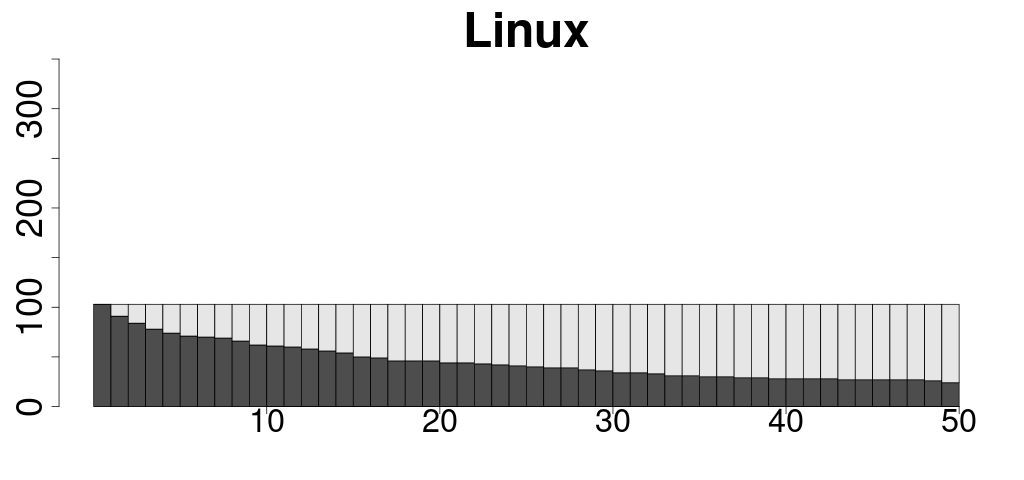}
\includegraphics[keepaspectratio=true, width=0.49\textwidth]{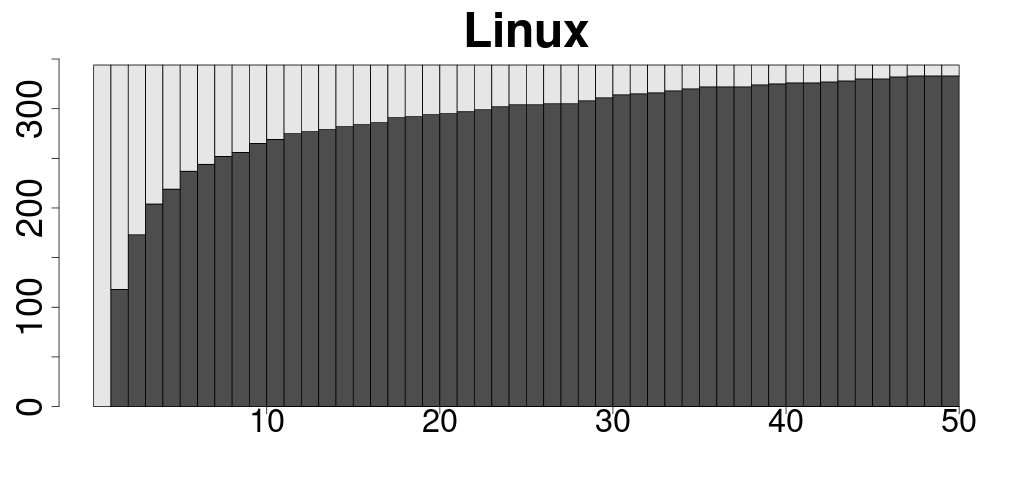}
\includegraphics[keepaspectratio=true,width=0.49\textwidth]{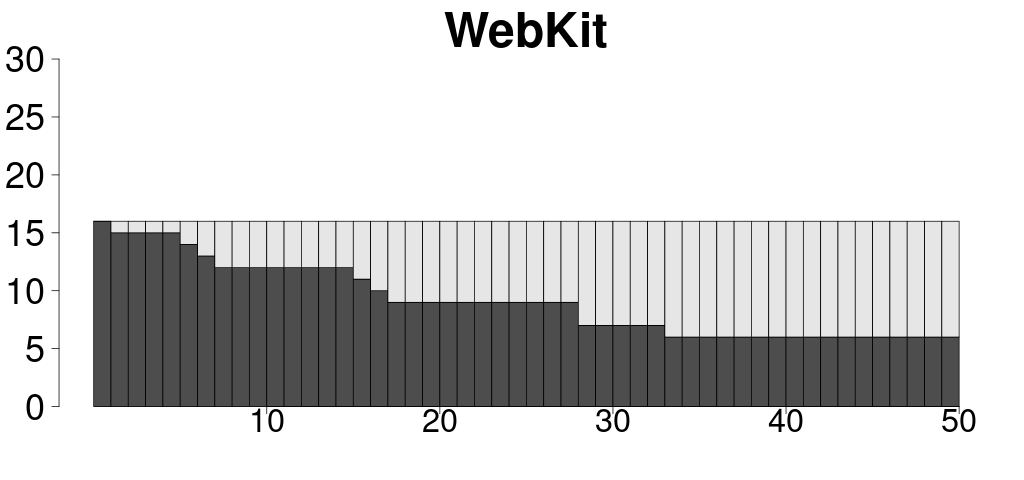}
\includegraphics[keepaspectratio=true, width=0.49\textwidth]{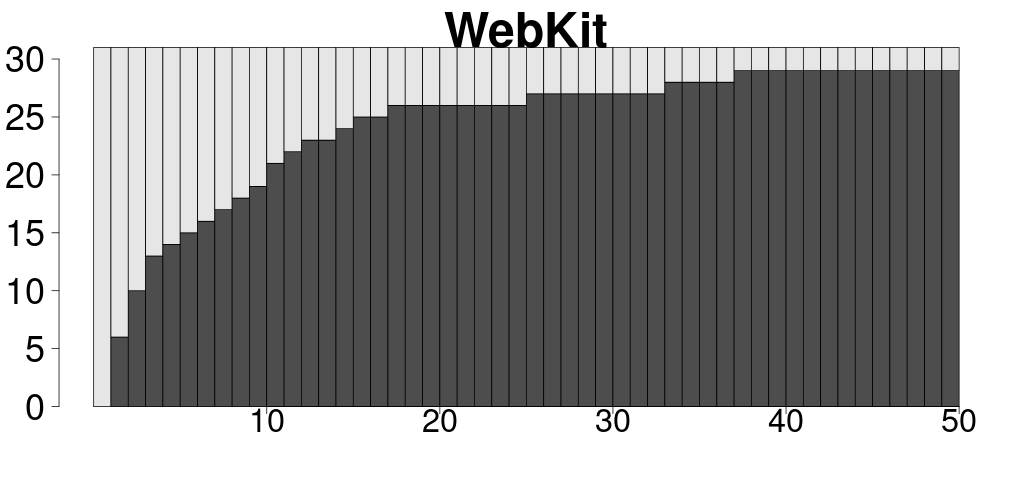}
\includegraphics[keepaspectratio=true,width=0.49\textwidth]{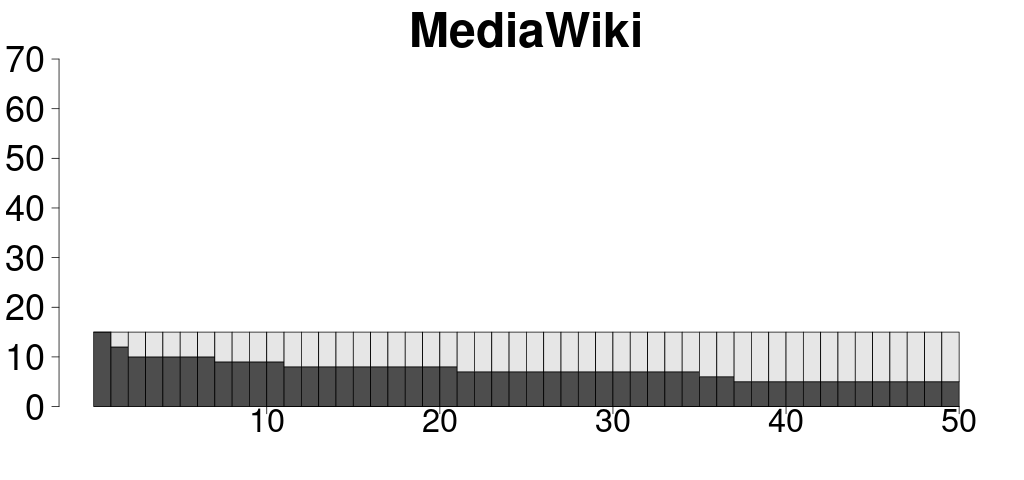}
\includegraphics[keepaspectratio=true, width=0.49\textwidth]{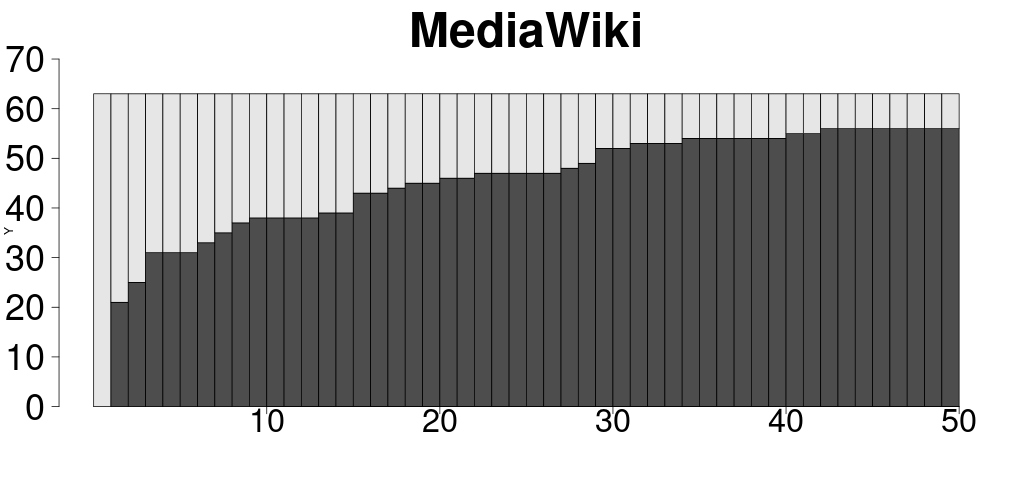}
\includegraphics[keepaspectratio=true,width=0.49\textwidth]{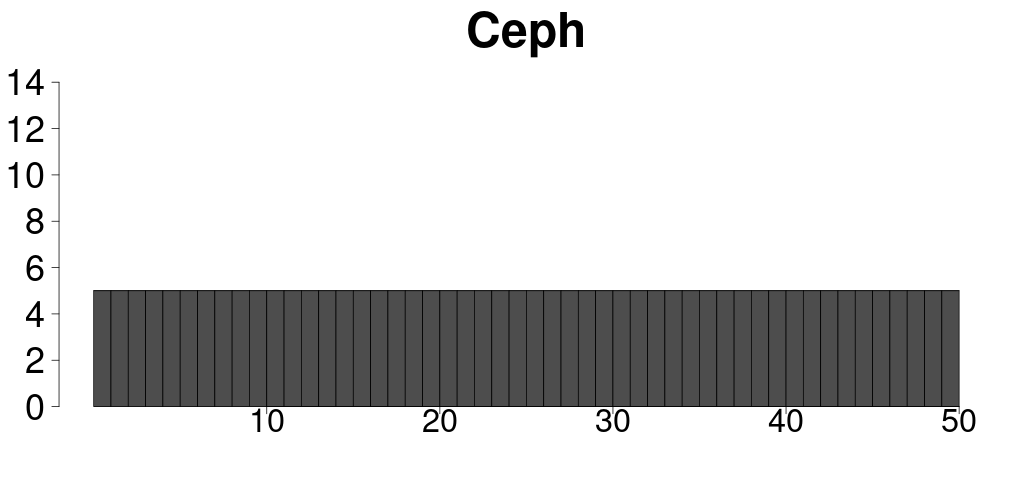}
\includegraphics[keepaspectratio=true, width=0.49\textwidth]{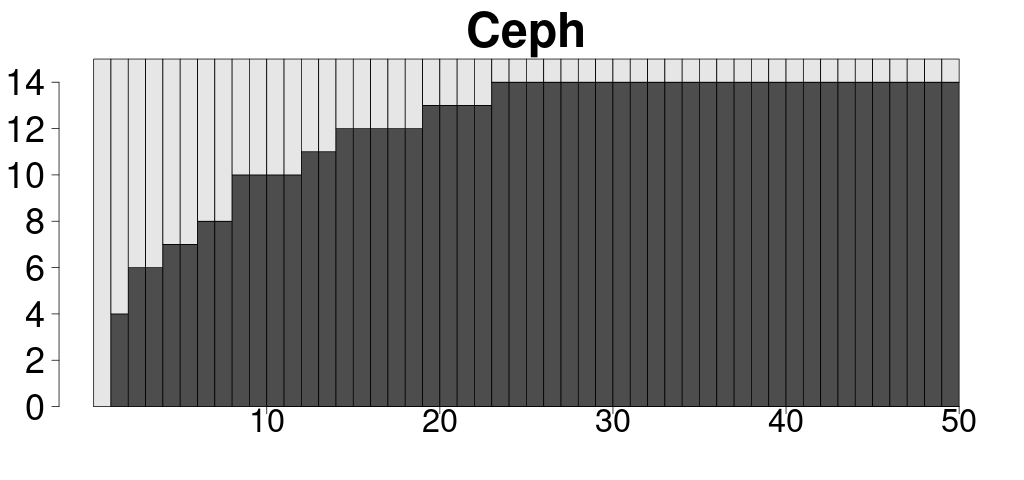}
\includegraphics[keepaspectratio=true,width=0.49\textwidth]{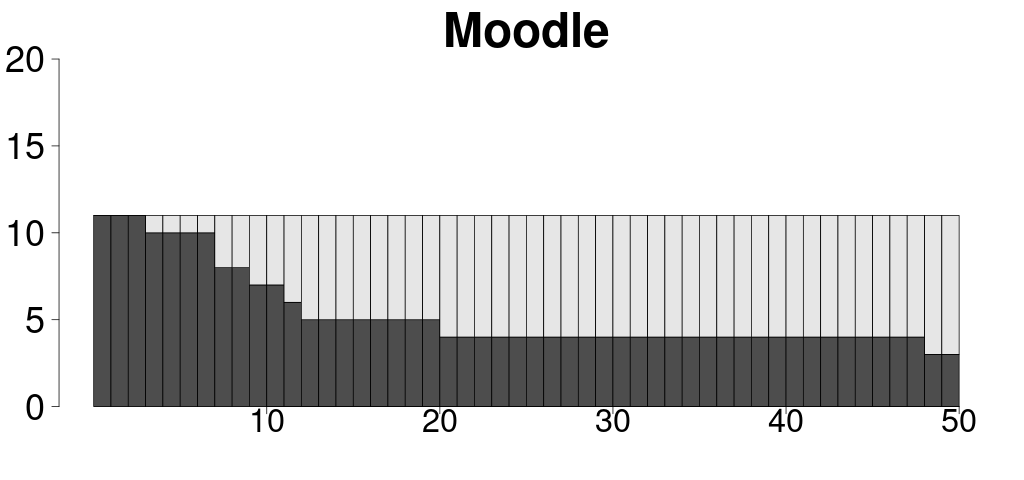}
\includegraphics[keepaspectratio=true, width=0.49\textwidth]{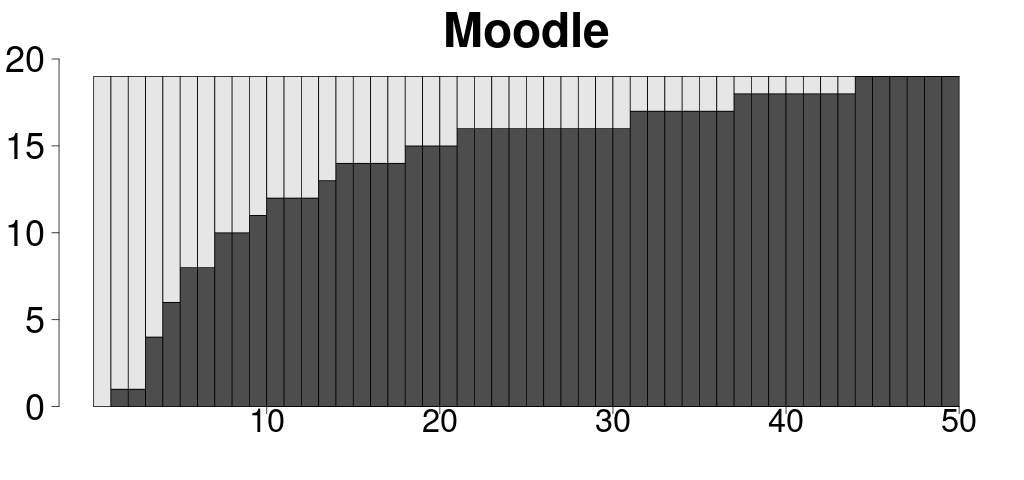}
	\caption{Correct identification of full-time (l) and non-full-time (r) developers depending on the threshold value (x-axis). The darker shade gives those developers that have been correctly identified, while the lighter one is for those that were not. The vertical axis has been kept the same for each project in order to ease visual comparison of the population of each group.}
	\label{fig:_tp_fn_tn_fp_others}
\end{center}
\end{figure*}

The approach that was presented for OpenStack was repeated for all the systems. 
In all the cases, when $\theta$ increases, the more precise it becomes to identify the non-full time developers, based to their responses to the questionnaires. Similarly, when $\theta$ grows, the accuracy at identifying full-time developers decreases for all but one project. 
Ceph (second from bottom plots in Figure~\ref{fig:_tp_fn_tn_fp_others}) is an expected variation to this trend: all its self-identified full-time developers have an extremely high number of commits in the six-month time-frame, higher than the 50 commits depicted in the figure.

\subsection{Compensation}
\label{sec:_compensation-further}

\begin{figure*}[ht!]
\begin{center}
\includegraphics[keepaspectratio=true,width=0.49\textwidth]{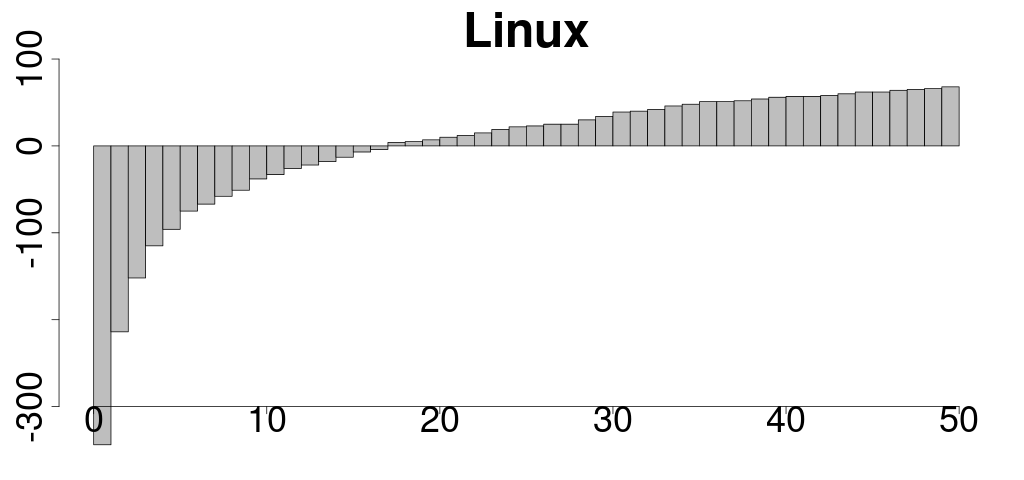}
\includegraphics[keepaspectratio=true,width=0.49\textwidth]{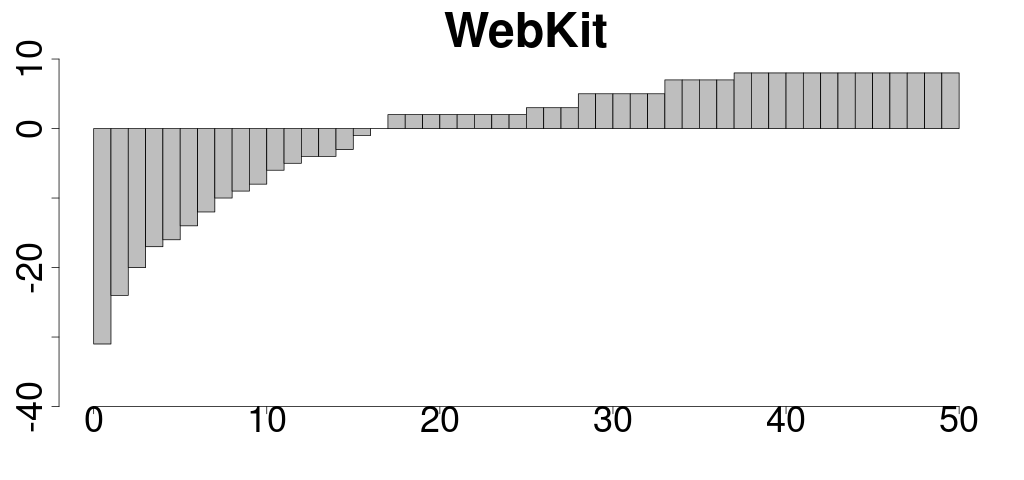}
\includegraphics[keepaspectratio=true,width=0.49\textwidth]{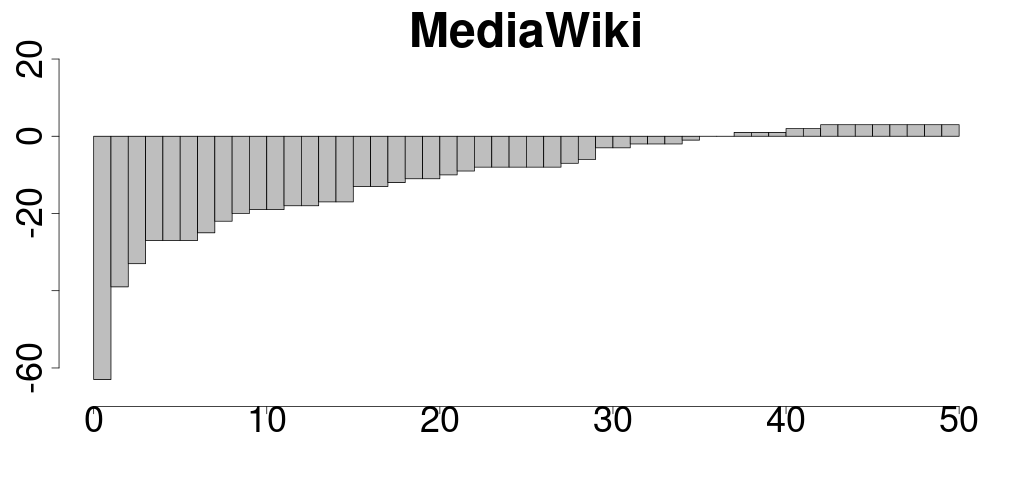}
\includegraphics[keepaspectratio=true,width=0.49\textwidth]{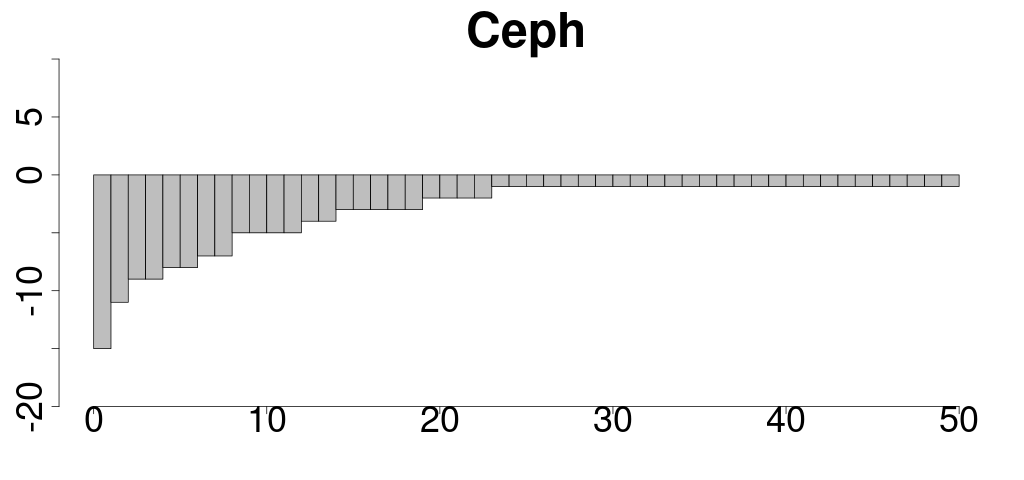}
\includegraphics[keepaspectratio=true,width=0.49\textwidth]{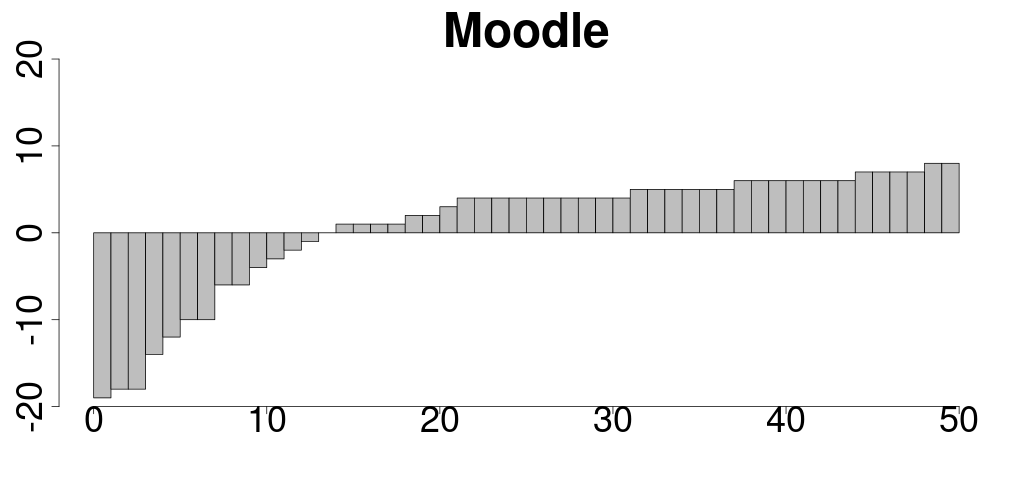}
\caption{Compensation between false positives and false negatives for values of $\theta$. Y-axis is given in number of developers; the x-axis corresponds to the $\theta$ value. Negatives values of the y axis indicate false positives that do not get compensated by false negatives, while positive values of the y axis indicate false negatives not compensated by false positives. Note that the magnitude of the vertical axis is different for each plot. 
}
\label{fig:_compensation_others}
\end{center}
\end{figure*}

Similarly to what achieved in Section~\ref{sec:compensation}, the compensation levels were evaluated to identify the minimum between false negatives and false positives. The plots of the compensation values, per project, are available in Figure~\ref{fig:_compensation_others}. Given a compensation plot, one can obtain the value of $\theta$ for each project, and more precisely where the compensation is closest to zero. In the extreme case of the Ceph project, there are no false negatives, and only false positives, therefore the compensation is negative for any threshold value.


\subsection{Goodness (and threshold values obtained)}
\label{sec:_goodness-further}

\begin{figure*}
\begin{center}
\includegraphics[keepaspectratio=true,width=0.86\textwidth]{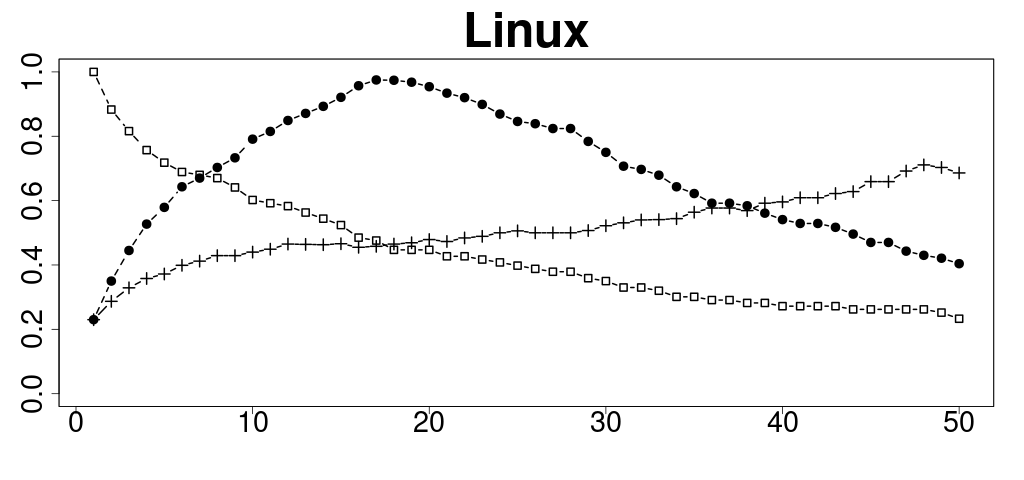}
\includegraphics[keepaspectratio=true,width=0.86\textwidth]{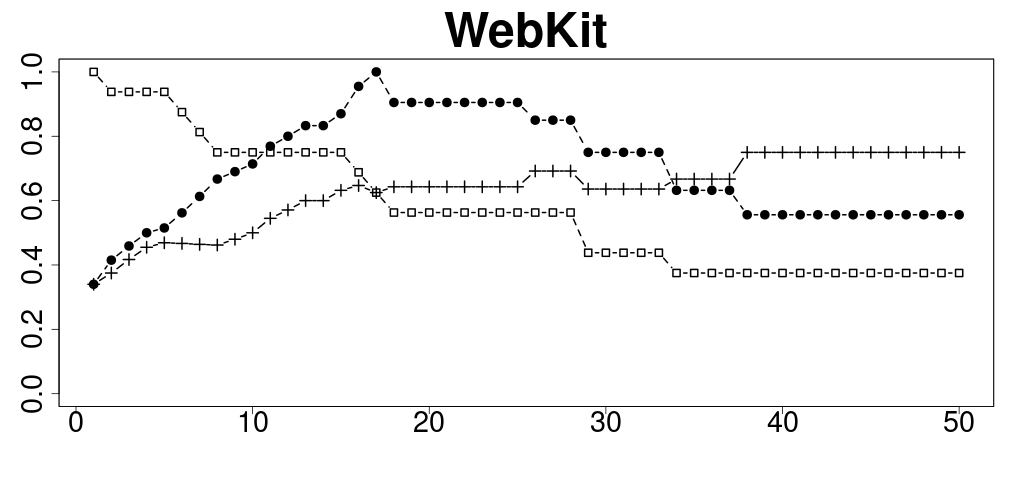}
\includegraphics[keepaspectratio=true,width=0.86\textwidth]{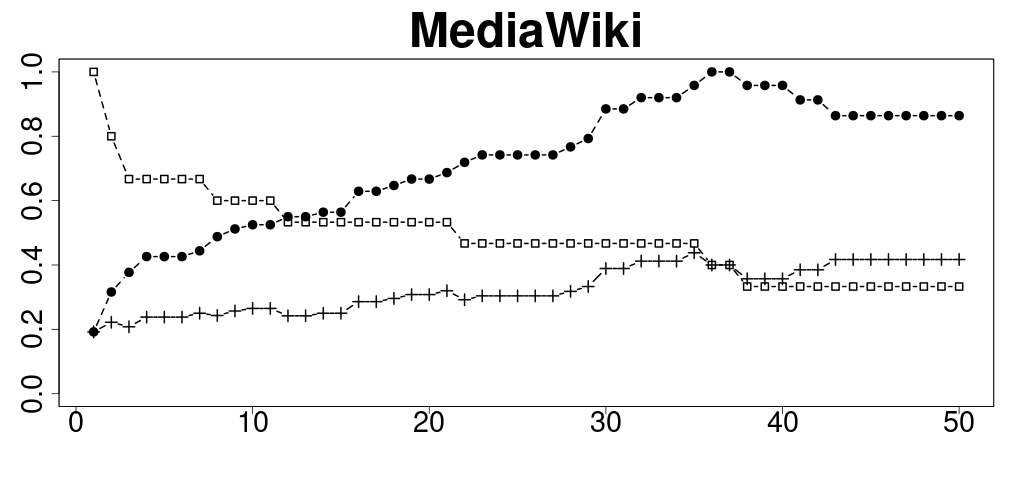}
\includegraphics[keepaspectratio=true,width=0.86\textwidth]{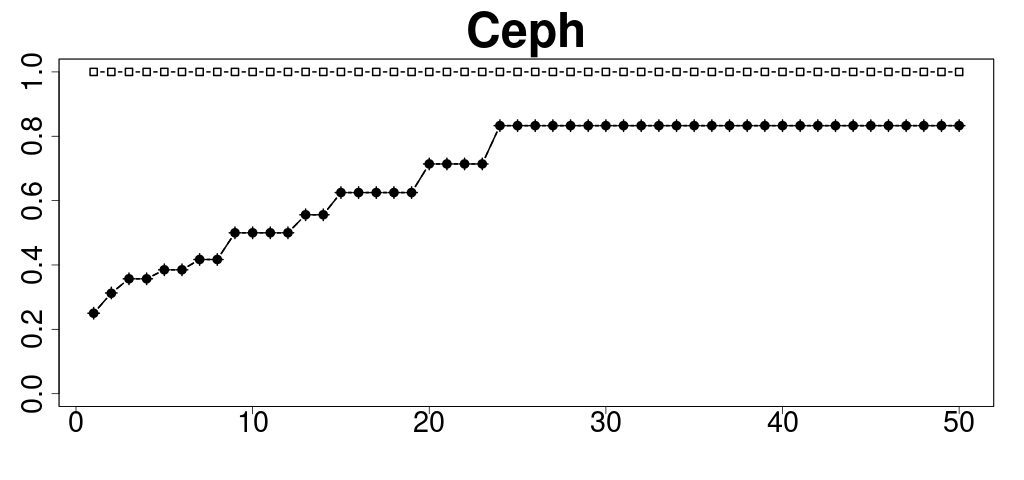}
\includegraphics[keepaspectratio=true,width=0.86\textwidth]{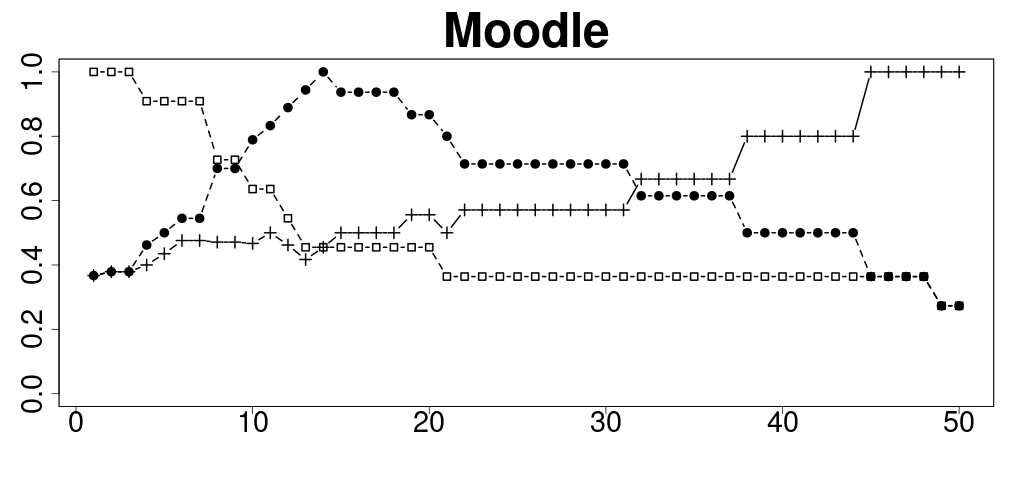}
	\caption{Values of precision, recall and goodness relative to $\theta$. Legend: Precision (\ding{59}), Recall (\ding{111}), Goodness (\ding{108}).  Note that in Ceph the values of goodness coincide with the ones of precision.
	}
	\label{fig:_relevance_others}
\end{center}
\end{figure*}

In Figure~\ref{fig:_relevance_others} we have plotted similar metrics to Figure~\ref{fig:_relevance_OpenStack} (i.e., precision, recall and goodness) for the rest of the analysed systems. The $\theta$ value for a project is selected in correspondence of the maximum value of the goodness: as demonstrated above, that threshold value minimizes the estimation error.


We collected the values of $\theta$, per project, in Table~\ref{tab:_threshold_gathered}. 
In the Discussion (in particular in Section~\ref{subsec:generalization}), we elaborate about why projects might have different values of $\theta$, and how this knowledge can be leveraged to extend our model to other projects, and without having to directly survey their developers.

\begin{table}[ht!]
\centering
\begin{tabular}{l|r}
Project   & $\theta$ \\\hline
OpenStack & 12          \\
Moodle    & 14          \\
Linux     & 17          \\
WebKit    & [16-19]          \\
Ceph      & 24          \\
MediaWiki & [36-37]          \\\hline
\end{tabular}
\caption{Values of $\theta$ for all the projects analysed}
\label{tab:_threshold_gathered}
\end{table}

\section{Discussion}
\label{sec:discussion}


In this section we discuss the limitations, implications and further possibilities of our research. We will start presenting the threats to validity (\ref{subsec:threats}), and then elaborate further on the sources of error of our model (\ref{subsec:error}). Finally, possible impact for practitioners and managers (\ref{subsec:repercussions}), the representativeness of our survey (\ref{subsec:representativeness}) and the possibilities of generalizing to other FOSS projects (\ref{subsec:generalization}) are presented and discussed.

\subsection{Threats to Validity}
\label{subsec:threats}
All empirical studies, such as this one, are subject to threats to validity. Here, we point the most relevant ones and discuss how to mitigate or control these threats if possible.

\textbf{Conceptual} Assumptions and assertions have been done. We have tried to make them explicit, as for instance the assumptions that a person-month equates to 40h/week. Some of them may require further support, but this does not invalidate the plausibility of the argument.

\textbf{Internal}
Having only considered one of the many sources of activity (and data) may result in our estimation model being not accurate.
Further work is expected to be of broad interest and supportive of the claims presented.


\textbf{External} Originally, OpenStack was chosen as the illustrative case study. We consider that this project is representative of many FOSS projects, especially those where many industry players cooperate. There may be certain practices that could only be found for OpenStack: for instance, their review process produces large (and few) commits, so values of $\theta$ should be selected differently for other FOSS projects. We have applied the model to other five FOSS projects, and the results are similar.

The model has been conceived to work well when participation follows a power-law and the number of full-time developers that can be easily identified because their activity is much higher then by the rest of developers -- else we could have many false negatives (i.e., many full-time developers not identified as such). What we have assumed is usually the case in FOSS development, but has not to be true for all FOSS projects. We have applied our model to large project (with hundreds of committers), but cannot ensure it will offer meaningful results for other FOSS projects, especially small ones.

The selection of the further 5 systems was based on convenience sampling: as said before, the communities of the projects selected were more easily reached, due to two of the authors of the being paper in direct contact with all of them. 
Although the sample might be unrepresentative, the systems were only used to deploy the model, not to show its absolute value for any other system. 
In any case, the generalization of the model to other FOSS projects remains to be investigated.



\textbf{Construct} A replication package\footnote{\url{http://gsyc.urjc.es/~grex/repro/2022-emse-effort-estimation}}, following the guidelines in~\cite{gonzalez2012reproducibility}, is offered to others to reproduce our findings. The replication package contains all scripts and public information, but not all the results of the surveys, as personal information was collected. Aggregated and anonymized information from the surveys is available.


It is important to note that
it is not possible to establish a unique \textit{ground truth} for the total effort made by developers in FOSS projects.
This is because FOSS projects do not typically make use of timesheets to track the actual effort of developers. Therefore, a `traditional' measure of effort is not clearly computable. 
The measures that we propose do not help in establishing what is the ground truth, but rather to minimise the error in the evaluation of the estimated effort based on actual input by the developers of the project.

A larger threat to construct validity was identified, based on the representativeness of the survey sample.
It should be, however, noted that for determining the best value of $\theta$, we do not need a representative sample of the \emph{whole} project population, but just a representative sample that allows us to effectively discriminate between full-time and non-full-time developers. We thus require basically data that is representative of (very) active developers to perform this analysis. With a very high probability, all the developers with a low activity (i.e., few commits) are non-full-time developers -- and our model will accurately label them as such.
We check if this affects our results in more detail in Section~\ref{subsec:representativeness}.


\subsection{On the sources of estimation error}
\label{subsec:error}

Our effort estimation model is based on the identification of full-time developers.
Once identified, full-timers are assigned the maximum possible effort (1PM every month): non-full-timers are only assigned a share of it, depending on the number of their commits and the $\theta$.
In this situation, the sources of error are the following:

  \begin{enumerate}
    \item Wrong identification of full-time developers (false negatives) or non-full-time developers (false positives).
    \item Wrong assignment of effort to non-full-timers.
  \end{enumerate}

Note that we assume that the assignment of effort to full-timers is error-free, as we argue that by definition the effort of a full-time developer is 1PM. We argue below (sections~\ref{subsec:fns} and~\ref{subsec:share}) why both sources of error do not pose a major threat to the validity of our model and its results.  


\subsubsection{Wrong identification}
\label{subsec:fns}
    
In our model false negatives (FN) refer to full-time developers that we have not identified as such. We have included a short comment in the threats to validity (Section~\ref{subsec:threats}), stating that results may not hold if there are many false negatives in a certain project. 

However, we consider it safe to assume that the number of false negatives will be small in the general case of FOSS development.
The reasons for this are as follows:

\begin{enumerate}
  \item The number of full-time developers should be (very) low compared to the amount of total participants in a FOSS project. This is because the distribution of participation is very skewed, as in a power-law or Rayleigh-type curve (as reported in the research literature~\cite{koch2008effort,sowe2008understanding}), in FOSS. So, in general, full-time developers will have profiles with high commit activity, well above the threshold value, that is, the lowest bound of activity for full-timers.

  \item A false negative might be compensated by a false positive. Some full-time developers might have a low activity, below $\theta$. When they are below $\theta$, compensation (i.e., non-full-time developers that are incorrectly identified as full-time (false positives)) softens this situation. In other words, because we have chosen $\theta$ where goodness is highest (i.e., the one that maximizes compensation), the number of FNs that are not compensated should be the lowest possible.

  \item If not compensated, the impact of the error is small, as an aggregate. Having a FN that is not compensated means that a developer is assigned a fraction of the effort depending on the number of commits and the threshold. So, if for OpenStack $\theta$ is 12 commits in a 6-month period, and we have a full-time developer with 6 commits in that period, we would have a false negative and this developer would be accounted with an effort of 6/12 * 6PM = 3 PM instead of with 6PM. This would imply an error of 100\% at the \emph{individual} level. But our estimation model is based on aggregation of all developers, so even if the error for one individual developer might be high, the fact that it is low for many other developers will pay off for the whole project.
\end{enumerate}


A similar reasoning applies to false positives (FP) in our model, i.e., to non-full-time developers that we have not identified as such. 
Even if the population of non-full-time developers was large, using the highest goodness value ensures highest compensation and, thus, the lowest aggregated number of FPs.
In this case, the individual error introduced would be because a non-full-time developer is assigned 6PMs in a period of six months, when the \emph{real} effort is less (e.g., 3PMs).
Again, the error at the \emph{individual} level might be large, but its impact on the whole project will be small.

For the six projects under study in this paper, the number of non-compensated FNs or FPs are: 1 developer for OpenStack, Linux and Ceph and 0 developers for WebKit, MediaWiki and Moodle. This can be seen from the Figures~\ref{fig:_compensation_OpenStack}~and~\ref{fig:_compensation_others}, 
This strengthens our argument: the impact of an incorrect identification of developers in our model is very low, even negligible in large FOSS projects.



\subsubsection{Wrong assignment of effort to non-full-timers}
\label{subsec:share}

Another source of error may come with the effort estimation assigned to non-full-time developers.
In our model we assign non-full-time developers the fraction of commits/threshold (i.e., if they did 2 commits in 6 months and $\theta$ is 12, as in OpenStack, then they will be assigned an effort of 1/6 * 6 = 1 PM for the 6 months). 

The rationale for this estimation is easy to understand: these developers are not directly estimated for their effort, but as its fraction to the minimum activity that is required to be considered as a full-time developer. 
In other words, if a non-full-time developer has authored 2 commits and we expect full-time developers to submit at least 12 commits ($\theta$ = 12), our model will assign this developer 1/6 of the minimum effort needed to be considered a full-time developer.

\subsection{Impact}
\label{subsec:repercussions}

Based on the models of FOSS development, the analysis that was presented, and its findings, can have a direct impact on the decisions taken by practitioners and managers. The estimation of past efforts, with the addition of an understanding of what is already implemented, what is missing, and what needs to be maintained, can help to plan for the future. 

If managers can estimate past effort, they can also evaluate the effort that is likely to be needed in the near future. This is particularly important for those projects where many developers, from different companies, contribute with very different levels of dedication, and where estimating past effort is very difficult. In those contexts, the impact of a company leaving the project can be estimated; in a similar way, the impact of a new company allocating its developers at different levels of effort (e.g, full-time, part-time, occasional, etc.). 
All these levels of developer engagement can be re-estimated periodically, so that managers can account from divergences from the plan.

In particular, from our experience the focus on the estimation of past effort in FOSS is important usually in two situations:

\begin{enumerate}
    \item When a company is evaluating the overall effort put in the development of a FOSS project, because they have a direct interest in assuming all or part of that cost. This is the case, for example, when the software is strategic to a company, and they want to evaluate how much effort was put in it in the past. This evaluation will be then used as an estimation of future effort, in the short and medium term, and considering to hire a part of the developing team: knowing the past effort gives also an evaluation of which fraction of the development effort they are hiring.

    \item When some FOSS is developed mainly by developers hired by companies, usually in the context of some Foundation or managing board, where companies involved decide about the resources they allocate to the project.
      In those case, having estimations about past efforts, and allocations of those efforts to companies (and to developers hired by those companies) is fundamental for the negotiations where new companies want to gain influence in the project, a seat in the board, or just negotiate how to move forward features that are important for them. We have specifically observed this kind of negotiations in projects such as Xen and OpenStack, and in some projects under the Linux Foundation umbrella.
\end{enumerate}

In addition to this, when companies participate to FOSS endeavours, they are fully aware of the effort produced by their own developers, and how they engage with the FOSS projects that they have an interest in. Having a better estimate of how much effort (overall) has been produced into a FOSS project would make a clearer case for a company's Return-on-Investment (RoI). As companies are usually aware of the effort they put into the project, they can compare this with the total effort obtaining a ratio of how much of this effort comes from the community. The returned effort is what a company gains for its stake over a FOSS, as contributed by the community around it. That gain would necessarily influence a company's strategic decision of investing its own developer efforts (or procure and pay for development efforts from other specialist companies) into the FOSS project, and at what level. Even more importantly, in case of other commercial enterprises participating to the same development, a company would be able to tailor and adapt their own input to the project, and based on other companies' behaviour.


\subsection{Representativeness of the survey sample}
\label{subsec:representativeness}

\begin{figure*}[ht!]
\begin{center}
\includegraphics[keepaspectratio=true,width=0.52\textwidth]{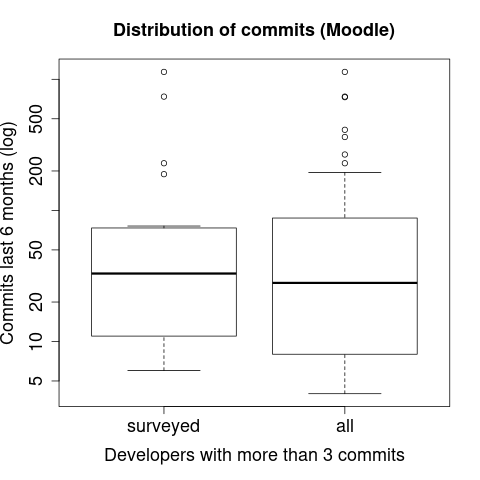}
	\caption{Boxplot with the activity (in number of commits during the last 6 months before the survey) for the active developers surveyed in the OpenStack project (left) and for all the active developers (right) in the last 6 months before the survey.}
	\label{fig:representativeness}
\end{center}
\end{figure*}

The rate of responses obtained in the questionnaires to the developers is around 25\% at best (in the case of the Moodle project). This means that most developers have not responded to the questionnaires, and the representativeness of the sample could be put in question. Limited to the OpenStack project, Figure~\ref{fig:representativeness} shows two box-plots with the analysis of the developers who responded to the survey (left) and all the active developers (right). The measured activity in number of commits considers the six months preceding the survey.

In order to check whether the two distributions of commits come from the same population, we applied the Kolmogorov-Smirnov's test, a non-parametric test that considering two samples, evaluates the null hypothesis $H_0$: \textit{are the two samples extracted from the same population?}. We considered various activity levels (e.g., developers committing 0 or more commits; developers committing 1 or more commits and so on) and tested their distribution against the overall active population of developers, at the same activity level. The significance level for each of those statistical tests was set to a standard $\alpha = 0.5$: Table~\ref{tab:wilcox} summarises the p-values (last column) along with other attributes.

\begin{figure*}[ht!]
\begin{center}
\includegraphics[keepaspectratio=true,width=0.32\textwidth]{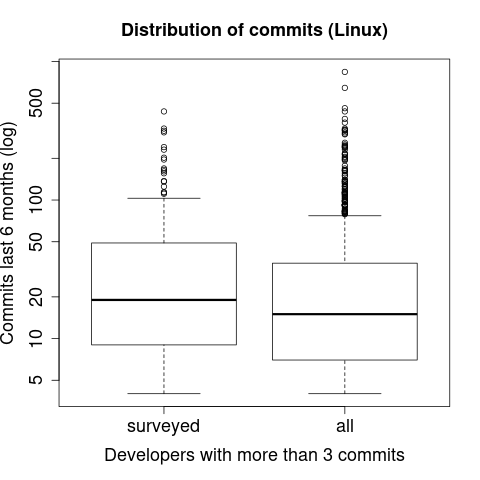}
\includegraphics[keepaspectratio=true,width=0.32\textwidth]{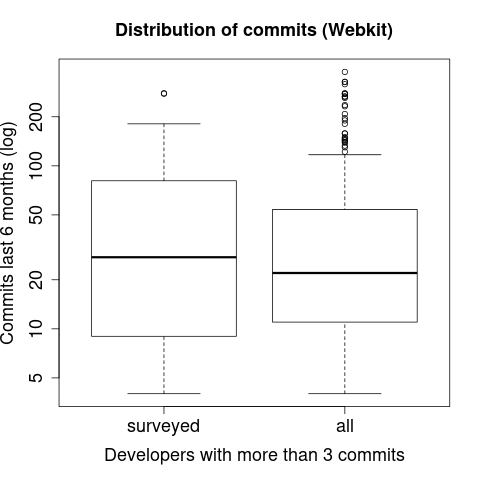}
\includegraphics[keepaspectratio=true,width=0.32\textwidth]{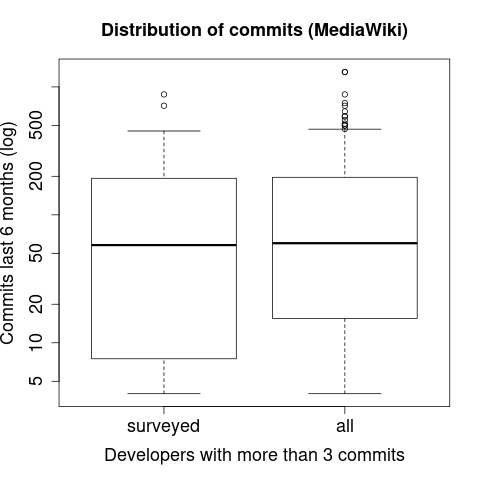}
\includegraphics[keepaspectratio=true,width=0.32\textwidth]{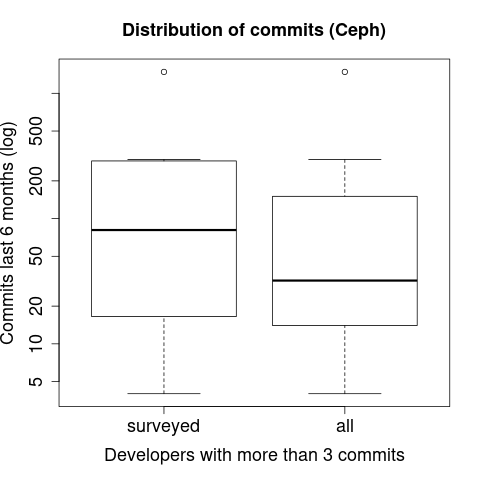}
\includegraphics[keepaspectratio=true,width=0.32\textwidth]{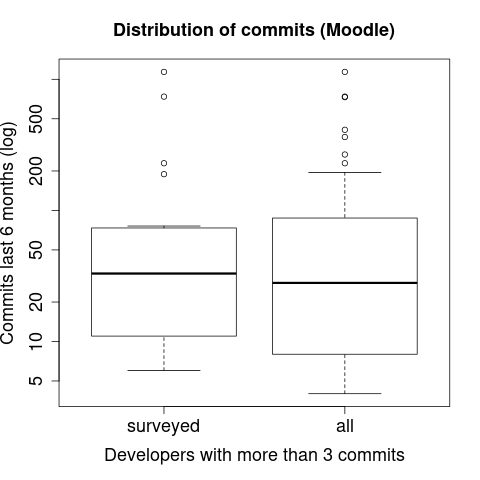}
	\caption{Boxplot with the activity (in number of commits during the last 6 months before the survey) for active surveyed developers and for all the active developers in the last 6 months before the survey.}
	\label{fig:representativeness_rest}
\end{center}
\end{figure*}

\begin{table*}[ht!]
\caption{Summary of population measures. Several populations have been selected,
depending on a minimum number of commits. \emph{D} and \emph{p-value} as given by the Two-sample Kolmogorov-Smirnov test.}
\center
\begin{tabular}{|c|c|cccccc|c|c|}
\hline
Commits & Population & Min. & 1st Q &  Median &   Mean & 3rd Q   &   Max. & D & p-value\\ \hline
$\geq$ 0 & all (1,626) &   0.00  &  0.00 &  1.00  & 9.62 &  6.00 & 491.00 & 0.203 & 0.0001 \\ 
  & survey (125) &  0.00  &  0.00 &   4.00  & 14.12  & 14.00 & 201.00 & & \\ \hline 
$\geq$ 1 & all (986) &  1.00  &  1.00 &   3.00  & 13.76  & 11.00 & 491.00 & 0.197 & 0.0028 \\
  & survey (92) &  1.00 &   1.00  &  5.00 &  16.35 &  19.00 & 201.00 & & \\ \hline
$\geq$ 2 & all (693) & 2.00 & 3.00 & 7.00 & 19.15 & 18.00 & 491.00 & 0.194 & 0.0128 \\
  & survey (74) &  2.00 &   5.00  &  12.00 &  23.61 &  27.00 & 201.00 & & \\ \hline 
$\geq$ 3 & all (563) & 3.00 & 5.00 & 9.00 & 23.11 & 22.00 & 491.00 & 0.196 & 0.0243 \\
  & survey (64) & 3.00 & 6.75 & 13.50 & 26.98 & 31.00 & 201.00 & & \\ \hline
$\geq$ 4 & all (490) & 4.00 & 6.00 & 11.00 & 26.11 & 26.00 & 491.00 & 0.136 & 0.2562 \\
  & survey (63) & 4.00 & 7.00 & 14.00 & 27.37 & 31.00 & 201.00 & &  \\ \hline    
$\geq$ 5 & all (427) & 5.00  &  7.00 &  13.00 &  29.37  & 30.00 & 491.00 & 0.121 & 0.4406 \\ 
  & survey (58) &  5.00 &   8.50 &  16.50 &  29.38 &  34.75 & 201.00 & & \\ \hline
$\geq$ 8 & all (314) & 8.00  &  12.00  &  19.00  &  37.77  & 41.00 & 491.00 & 0.119 & 0.6229 \\ 
  & survey (46) &   8.00 &  13.00  & 22.00  & 35.57 &  42.75 & 201.00 & & \\ \hline
$\geq$ 11 & all (256) &  11.00 &  14.75 &  24.00 &  44.34 &  52.00 & 491.00 & 0.084 & 0.9631 \\ 
   & survey (41) &  11.00 &  14.00 &  26.00 & 38.83  & 50.00 & 201.00 & & \\ \hline
\end{tabular}
\label{tab:wilcox}
\end{table*}

As visible from Table~\ref{tab:wilcox}, we reject the base $H_0$ at all activity levels up to 4 commits per period: hence, between 1 and 4 commits, we \textit{can} reject the hypothesis that the surveyed developers represent the overall population. With higher commit activities (5 commits or more), the $H_0$ cannot be rejected at the $\alpha=0.05$ level: from 5 commits up, we \textit{cannot} reject the hypothesis that the surveyed developers represent the overall population. As the values of activity increase, the surveyed population becomes more representative of the project: since our model is based on activity and the classification is performed only on active developers, these results give a stronger support to our model.

The same analysis was extended to the other sampled systems, and summarised in the boxplots of Figure~\ref{fig:representativeness_rest}. From top left, the activity of the responding developers (in the 6 months preceding the questionnaire) for the Linux, WebKit, MediaWiki, Ceph and Moodle systems was compared with the activity of all developers in each project. Similarly to OpenStack, we \textit{cannot} reject the hypothesis that the surveyed developers of those projects represent the overall population, for activities of more than 3 commits.


\subsection{Generalization to other projects}
\label{subsec:generalization}

If a project wants to know its own $\theta$ with accuracy, the amount of data that they need is limited and is easy to gather.
By polling their developers for their full-time/part-time status, they could use our model to find their particular $\theta$.
Some FOSS projects regularly survey their developers for knowing their community better, including personal, academic, working, community and other matters.
So, for instance, since 2011 OpenStack performs a yearly developer survey\footnote{https://insights.stackoverflow.com/survey/}.
Gathering information for our model would just require two questions to be added to those surveys (e.g., `\textit{do you consider yourself to be a full-time developer}' and `\textit{how many commits have you performed in the last 6 months}') .

We think, however, that the value of $\theta$ is dependent on intrinsic properties of the project, and that a FOSS project can use the model without having to perform a developers survey, but to use a $\theta$ from a similar project.
The results of applying our model to the 6 FOSS projects in this study make us think that $\theta$ depends on the process that the project follows.
Thus, depending on the project's practices, commits may require more or less effort to be approved before being merged, depending on whether a review process is in practice. 
If that is the case, code is not formally committed to the repository until it has been through extensive review, usually including several revisions. 
In addition, and particularly in those projects that have more strict code review practices, projects require to \textit{squash} all commits into a single one once a change is accepted; then, this only commit can be pushed to the repository.
This has as a consequence that commits are larger and more costly (in time and effort) for those projects, than for others that do not follow this practice. 

This can be confirmed from the projects that we have studied in this paper.
The values of $\theta$ shown in Table~\ref{tab:_threshold_gathered} are similar for those projects that have strict code reviewing practices, such as OpenStack ($\theta$=12), Moodle (14), Linux (17) and WebKit (16-19). 
All of them use Gerrit to support code review.
Ceph (24) and MediaWiki (36-37) have higher numbers of $\theta$, because their code review practices are \emph{lighter} (i.e., less demanding) and commit squashing is not that strictly followed as in the former projects.
For instance, Ceph uses GitHub's pull-request mechanism, which requires just one approval from a maintainer, while for OpenStack changes require to be approved by at least two maintainers.

Projects that do not use commit squashing, could use `active days' instead of `commits' in the model. 
That way, all commits during the same day would be considered as a contribution, mitigating the impact of different commit behaviours of committers~\cite{kolassa2013empirical}.

\section{Conclusion and Future Work}
\label{sec:conclusions}

This paper has tackled two challenges. The first is how to design a simple, but sound estimation model to measure the effort provided in a sparse, distributed and uneven development scenario, like the FOSS one. The second challenge is how to design the model so that it offers not only a reasonable prediction, but also credible. 

In order to maximise the \textit{simplicity} of our estimation model for FOSS development, we only discriminate between two types of developers: \emph{full-timers} and the rest. For harnessing the \textit{credibility} of our model, we have obtained feedback data from over a thousand developers, who work at different levels across six large FOSS projects.

The model establishes a threshold $\theta$ that separates the level of activity of \emph{full-timers} from the rest of developers. The value of $\theta$ has been optimised with the developers' responses: thanks to their feedback, we have achieved a much more realistic separation of developer types, minimizing the estimation error. Using $\theta$ in our model, the estimation of the overall effort results in a simple calculation, and just using two developer types.

We conjecture that the model being dependent on this value of $\theta$ is what allows the model to be useful for different projects: the relationship of commits to effort (which is at the core of the model) may be very different from project to project, but we have shown that it can be captured with this single parameter.

Using 6 large FOSS projects as case studies, we have shown how the model can be applied and fine-tuned. 
Although further research on this is needed, the results obtained make us hypothesize that the value of $\theta$ depends on the development practices -- basically, how strict code reviewing practices are and if commit squashing is frequently used when merging changes into the source code.
If so, the future applicability of our model would not require to survey developers, as it would suffice to use a value of $\theta$ obtained from projects that follow similar practices.

We envisage to expand this study by 1) studying other FOSS projects to ascertain if our method is applicable in general, and if so, to what extent; 2) performing a scientific experiment to obtain margins of error for the estimation of error for non-full-time developers; 3) comparing our results with the ones provided by traditional software estimation models used in industry, such as COCOMO; 4) after quantifying the effort required in a FOSS project, establishing whether it is more profitable for a prospective adopting company to redo (``make'') their own system, or to invest (``buy'') in the existing FOSS system as discussed in~\cite{asundi2005need}; and 5) comparing our approach with previous effort estimation techniques for FOSS projects, as the one proposed in~\cite{capra2007economics} based on the measure of entropy to calculate maintenance costs.

\begin{acknowledgements}
We want to express our gratitude to Bitergia\footnote{\url{http://bitergia.com/}} for the support they have provided when questions have arisen. 
We acknowledge the support of the Government of Spain through the “BugBirth” project (RTI2018-101963-B-100). We also acknowledge the work by Carlos Cervigón on an earlier version of the manuscript.

\end{acknowledgements}


\bibliographystyle{abbrv}      
\bibliography{effort-estimation.bib}   

\begin{thebibliography}{10}

\bibitem{abdelmoez2012bug}
W.~Abdelmoez, M.~Kholief, and F.~M. Elsalmy.
\newblock Bug fix-time prediction model using na{\"\i}ve bayes classifier.
\newblock In {\em 2012 22nd International Conference on Computer Theory and
  Applications (ICCTA)}, pages 167--172. IEEE, 2012.

\bibitem{abran20163}
A.~Abran, J.-M. Desharnais, and F.~Aziz.
\newblock 3.5 measurement convertibility—from function points to cosmic ffp.
\newblock {\em Cosmic Function Points: Theory and Advanced Practices}, page
  214, 2016.

\bibitem{agrawal2018we}
A.~Agrawal, A.~Rahman, R.~Krishna, A.~Sobran, and T.~Menzies.
\newblock We don't need another hero? the impact of" heroes" on software
  development.
\newblock In {\em Proceedings of the 40th International Conference on Software
  Engineering: Software Engineering in Practice}, pages 245--253, 2018.

\bibitem{ahsan2009program}
S.~N. Ahsan, J.~Ferzund, and F.~Wotawa.
\newblock Program file bug fix effort estimation using machine learning methods
  for oss.
\newblock In {\em SEKE}, pages 129--134, 2009.

\bibitem{alomari2015slicing}
H.~Alomari.
\newblock A slicing-based effort estimation approach for open-source software
  projects.
\newblock {\em International Journal of Advance Computational Engineering and
  Networking (IJACEN)}, 3(8):1--7, 2015.

\bibitem{amor2006effort}
J.~J. Amor, G.~Robles, and J.~M. Gonzalez-Barahona.
\newblock Effort estimation by characterizing developer activity.
\newblock In {\em Proceedings of the 2006 international workshop on Economics
  driven software engineering research}, pages 3--6. ACM, 2006.

\bibitem{anbalagan2009predicting}
P.~Anbalagan and M.~Vouk.
\newblock On predicting the time taken to correct bug reports in {O}pen
  {S}ource projects.
\newblock In {\em Software Maintenance, 2009. ICSM 2009. IEEE International
  Conference on}, pages 523--526. IEEE, 2009.

\bibitem{asundi2005need}
J.~Asundi.
\newblock The need for effort estimation models for open source software
  projects.
\newblock {\em ACM SIGSOFT Software Engineering Notes}, 30(4):1--3, 2005.

\bibitem{barry1981software}
B.~Boehm.
\newblock Software engineering economics, 1981.

\bibitem{boehm2000software}
B.~W. Boehm, R.~Madachy, B.~Steece, et~al.
\newblock {\em Software Cost Estimation with {COCOMO} {II} with {CDROM}}.
\newblock Prentice Hall PTR, 2000.

\bibitem{capiluppi2013effort}
A.~Capiluppi and D.~Izquierdo-Cort{\'a}zar.
\newblock Effort estimation of {FLOSS} projects: a study of the {L}inux kernel.
\newblock {\em Empirical Software Engineering}, 18(1):60--88, 2013.

\bibitem{capiluppi2007cathedral}
A.~Capiluppi and M.~Michlmayr.
\newblock From the cathedral to the bazaar: An empirical study of the lifecycle
  of volunteer community projects.
\newblock In {\em IFIP International Conference on Open Source Systems}, pages
  31--44. Springer, 2007.

\bibitem{capra2007economics}
E.~Capra, C.~Francalanci, and F.~Merlo.
\newblock The economics of open source software: an empirical analysis of
  maintenance costs.
\newblock In {\em Software Maintenance, 2007. ICSM 2007. IEEE International
  Conference on}, pages 395--404. IEEE, 2007.

\bibitem{capra2008empirical}
E.~Capra, C.~Francalanci, and F.~Merlo.
\newblock An empirical study on the relationship between software design
  quality, development effort and governance in {O}pen {S}ource {P}rojects.
\newblock {\em Software Engineering, IEEE Transactions on}, 34(6):765--782,
  2008.

\bibitem{capra2010economics}
E.~Capra, C.~Francalanci, and F.~Merlo.
\newblock The economics of community open source software projects: an
  empirical analysis of maintenance effort.
\newblock {\em Advances in Software Engineering}, 2010, 2010.

\bibitem{crowston2005social}
K.~Crowston and J.~Howison.
\newblock The social structure of free and open source software development.
\newblock {\em First Monday}, 10(2), 2005.

\bibitem{duenas2018perceval}
S.~Due{\~n}as, V.~Cosentino, G.~Robles, and J.~M. Gonzalez-Barahona.
\newblock Perceval: Software project data at your will.
\newblock In {\em Proceedings of the 40th International Conference on Software
  Engineering: Companion Proceeedings}, pages 1--4, 2018.

\bibitem{dumke2016cosmic}
R.~Dumke and A.~Abran.
\newblock {\em COSMIC Function Points: Theory and Advanced Practices}.
\newblock Auerbach Publications, 2016.

\bibitem{fernandez2009does}
J.~Fernandez-Ramil, D.~Izquierdo-Cortazar, and T.~Mens.
\newblock What does it take to develop a million lines of {O}pen {S}ource code?
\newblock In {\em Open Source Ecosystems: Diverse Communities Interacting},
  pages 170--184. Springer, 2009.

\bibitem{fitzgerald2006transformation}
B.~Fitzgerald.
\newblock The transformation of {O}pen {S}ource {S}oftware.
\newblock {\em Mis Quarterly}, pages 587--598, 2006.

\bibitem{gonzalez2012reproducibility}
J.~M. Gonz{\'a}lez-Barahona and G.~Robles.
\newblock On the reproducibility of empirical software engineering studies
  based on data retrieved from development repositories.
\newblock {\em Empirical Software Engineering}, 17(1-2):75--89, 2012.

\bibitem{honel2018changeset}
S.~H{\"o}nel, M.~Ericsson, W.~L{\"o}we, and A.~Wingkvist.
\newblock A changeset-based approach to assess source code density and
  developer efficacy.
\newblock In {\em Proceedings of the 40th International Conference on Software
  Engineering: Companion Proceeedings}, pages 220--221, 2018.

\bibitem{hou2014empirical}
Q.~Hou, Y.~Ma, J.~Chen, and Y.~Xu.
\newblock An empirical study on inter-commit times in svn.
\newblock In {\em SEKE}, pages 132--137, 2014.

\bibitem{jorgensen2007systematic}
M.~Jorgensen and M.~Shepperd.
\newblock A systematic review of software development cost estimation studies.
\newblock {\em Software Engineering, IEEE Transactions on}, 33(1):33--53, 2007.

\bibitem{kalliamvakou2014promises}
E.~Kalliamvakou, G.~Gousios, K.~Blincoe, L.~Singer, D.~M. German, and
  D.~Damian.
\newblock The promises and perils of mining github.
\newblock In {\em Proceedings of the 11th working conference on mining software
  repositories}, pages 92--101, 2014.

\bibitem{kalliamvakou2009measuring}
E.~Kalliamvakou, G.~Gousios, D.~Spinellis, and N.~Pouloudi.
\newblock Measuring developer contribution from software repository data.
\newblock {\em MCIS}, 2009:4th, 2009.

\bibitem{koch2004profiling}
S.~Koch.
\newblock Profiling an open source project ecology and its programmers.
\newblock {\em Electronic Markets}, 14(2):77--88, 2004.

\bibitem{koch2008effort}
S.~Koch.
\newblock Effort modeling and programmer participation in open source software
  projects.
\newblock {\em Information Economics and Policy}, 20(4):345--355, 2008.

\bibitem{koch2002effort}
S.~Koch and G.~Schneider.
\newblock Effort, co-operation and co-ordination in an open source software
  project: {GNOME}.
\newblock {\em Information Systems Journal}, 12(1):27--42, 2002.

\bibitem{kolassa2013empirical}
C.~Kolassa, D.~Riehle, and M.~A. Salim.
\newblock The empirical commit frequency distribution of open source projects.
\newblock In {\em Proceedings of the 9th International Symposium on Open
  Collaboration}, pages 1--8, 2013.

\bibitem{kolassa2013model}
C.~Kolassa, D.~Riehle, and M.~A. Salim.
\newblock A model of the commit size distribution of open source.
\newblock In {\em International Conference on Current Trends in Theory and
  Practice of Computer Science}, pages 52--66. Springer, 2013.

\bibitem{kononenko2018studying}
O.~Kononenko, T.~Rose, O.~Baysal, M.~Godfrey, D.~Theisen, and B.~De~Water.
\newblock Studying pull request merges: a case study of shopify's active
  merchant.
\newblock In {\em Proceedings of the 40th International Conference on Software
  Engineering: Software Engineering in Practice}, pages 124--133, 2018.

\bibitem{kouters2012s}
E.~Kouters, B.~Vasilescu, A.~Serebrenik, and M.~G. van~den Brand.
\newblock Who's who in {GNOME}: Using {LSA} to merge software repository
  identities.
\newblock In {\em Software Maintenance (ICSM), 2012 28th IEEE International
  Conference on}, pages 592--595. IEEE, 2012.

\bibitem{lerner2002some}
J.~Lerner and J.~Tirole.
\newblock Some simple economics of open source.
\newblock {\em The journal of industrial economics}, 50(2):197--234, 2002.

\bibitem{ma2014dynamics}
Y.~Ma, Y.~Wu, and Y.~Xu.
\newblock Dynamics of open-source software developer's commit behavior: an
  empirical investigation of subversion.
\newblock In {\em Proceedings of the 29th Annual ACM Symposium on Applied
  Computing}, pages 1171--1173, 2014.

\bibitem{malhotra2020using}
R.~Malhotra and K.~Lata.
\newblock Using ensembles for class-imbalance problem to predict
  maintainability of open source software.
\newblock {\em International Journal of Reliability, Quality and Safety
  Engineering}, page 2040011, 2020.

\bibitem{mi2016empirical}
Q.~Mi and J.~Keung.
\newblock An empirical analysis of reopened bugs based on open source projects.
\newblock In {\em Proceedings of the 20th International Conference on
  Evaluation and Assessment in Software Engineering}, pages 1--10, 2016.

\bibitem{michlmayr2015and}
M.~Michlmayr, B.~Fitzgerald, and K.-J. Stol.
\newblock Why and how should open source projects adopt time-based releases?
\newblock {\em IEEE Software}, 32(2):55--63, 2015.

\bibitem{mockus2002two}
A.~Mockus, R.~T. Fielding, and J.~D. Herbsleb.
\newblock Two case studies of open source software development: Apache and
  mozilla.
\newblock {\em ACM Transactions on Software Engineering and Methodology
  (TOSEM)}, 11(3):309--346, 2002.

\bibitem{mockus2000identifying}
A.~Mockus and L.~G. Votta.
\newblock Identifying reasons for software changes using historic databases.
\newblock In {\em Software Maintenance, 2000. Proceedings. International
  Conference on}, pages 120--130. IEEE, 2000.

\bibitem{moulla2013cocomo}
D.~Moulla and Kolyang.
\newblock {COCOMO} model for software based on open source: Application to the
  adaptation of triade to the university system.
\newblock {\em International Journal on Computer Science and Engineering
  (IJCSE)}, 5(6):522--527, 2013.

\bibitem{moulla2014application}
D.~K. Moulla, I.~Damakoa, and D.~T. Kolyang.
\newblock Application of function points to software based on open source: A
  case study.
\newblock In {\em 2014 Joint Conference of the International Workshop on
  Software Measurement and the International Conference on Software Process and
  Product Measurement}, pages 191--195. IEEE, 2014.

\bibitem{porru2016estimating}
S.~Porru, A.~Murgia, S.~Demeyer, M.~Marchesi, and R.~Tonelli.
\newblock Estimating story points from issue reports.
\newblock In {\em Proceedings of the The 12th International Conference on
  Predictive Models and Data Analytics in Software Engineering}, pages 1--10,
  2016.

\bibitem{riehle2014paid}
D.~Riehle, P.~Riemer, C.~Kolassa, and M.~Schmidt.
\newblock Paid vs. volunteer work in open source.
\newblock In {\em 2014 47th Hawaii International Conference on System
  Sciences}, pages 3286--3295. IEEE, 2014.

\bibitem{robles2005developer}
G.~Robles and J.~M. Gonzalez-Barahona.
\newblock Developer identification methods for integrated data from various
  sources.
\newblock {\em ACM SIGSOFT Software Engineering Notes}, 30(4):1--5, 2005.

\bibitem{robles2014estimating}
G.~Robles, J.~M. Gonz{\'a}lez-Barahona, C.~Cervig{\'o}n, A.~Capiluppi, and
  D.~Izquierdo-Cort{\'a}zar.
\newblock Estimating development effort in free/open source software projects
  by mining software repositories: a case study of openstack.
\newblock In {\em Proceedings of the 11th Working Conference on Mining Software
  Repositories}, pages 222--231. ACM, 2014.

\bibitem{robles2004remote}
G.~Robles, S.~Koch, J.~M. GonZ{\'A}lEZ-BARAHonA, and J.~Carlos.
\newblock Remote analysis and measurement of libre software systems by means of
  the cvsanaly tool.
\newblock In {\em Proceedings of the 2nd ICSE Workshop on Remote Analysis and
  Measurement of Software Systems (RAMSS)}, pages 51--56. IET, 2004.

\bibitem{shah2006motivation}
S.~K. Shah.
\newblock Motivation, governance, and the viability of hybrid forms in open
  source software development.
\newblock {\em Management science}, 52(7):1000--1014, 2006.

\bibitem{sowe2008understanding}
S.~K. Sowe, I.~Stamelos, and L.~Angelis.
\newblock Understanding knowledge sharing activities in free/open source
  software projects: An empirical study.
\newblock {\em Journal of Systems and Software}, 81(3):431--446, 2008.

\bibitem{steinmacher2015social}
I.~Steinmacher, T.~Conte, M.~A. Gerosa, and D.~Redmiles.
\newblock Social barriers faced by newcomers placing their first contribution
  in open source software projects.
\newblock In {\em Proceedings of the 18th ACM conference on Computer supported
  cooperative work \& social computing}, pages 1379--1392, 2015.

\bibitem{thung2016automatic}
F.~Thung.
\newblock Automatic prediction of bug fixing effort measured by code churn
  size.
\newblock In {\em Proceedings of the 5th International Workshop on Software
  Mining}, pages 18--23, 2016.

\bibitem{von2003community}
G.~Von~Krogh, S.~Spaeth, and K.~R. Lakhani.
\newblock Community, joining, and specialization in open source software
  innovation: a case study.
\newblock {\em Research policy}, 32(7):1217--1241, 2003.

\bibitem{wiese2016mailing}
I.~S. Wiese, J.~T. da~Silva, I.~Steinmacher, C.~Treude, and M.~A. Gerosa.
\newblock Who is who in the mailing list? comparing six disambiguation
  heuristics to identify multiple addresses of a participant.
\newblock In {\em 2016 IEEE international conference on software maintenance
  and evolution (ICSME)}, pages 345--355. IEEE, 2016.

\bibitem{wu2016maintenance}
H.~Wu, L.~Shi, C.~Chen, Q.~Wang, and B.~Boehm.
\newblock Maintenance effort estimation for open source software: A systematic
  literature review.
\newblock In {\em Software Maintenance and Evolution (ICSME), 2016 IEEE
  International Conference on}, pages 32--43. IEEE, 2016.

\bibitem{yang2016empirical}
Y.~Yang, M.~Harman, J.~Krinke, S.~Islam, D.~Binkley, Y.~Zhou, and B.~Xu.
\newblock An empirical study on dependence clusters for effort-aware
  fault-proneness prediction.
\newblock In {\em 2016 31st IEEE/ACM International Conference on Automated
  Software Engineering (ASE)}, pages 296--307. IEEE, 2016.

\bibitem{yu2006indirectly}
L.~Yu.
\newblock Indirectly predicting the maintenance effort of {O}pen-{S}ource
  {S}oftware.
\newblock {\em Journal of Software Maintenance and Evolution: Research and
  Practice}, 18(5):311--332, 2006.

\bibitem{zhao2016discussions}
Y.~Zhao, F.~Zhang, E.~Shihab, Y.~Zou, and A.~E. Hassan.
\newblock How are discussions associated with bug reworking? an empirical study
  on open source projects.
\newblock In {\em Proceedings of the 10th ACM/IEEE International Symposium on
  Empirical Software Engineering and Measurement}, pages 1--10, 2016.

\end{thebibliography}

\end{document}